\documentclass[prd,onecolumn,superscriptaddress,floatfix,nopacs,preprintnumbers,nofootinbib]{revtex4}
\usepackage[utf8]{inputenc}

\usepackage{graphicx}
\usepackage[normalem]{ulem}
\usepackage{xcolor}

\usepackage{mathrsfs}
\usepackage{amssymb,bm}
\usepackage{amsmath}
\usepackage{mathtools}
\usepackage{physics}
\usepackage{slashed}
\usepackage{verbatim}

\newcommand{\nc}{N_\mathrm{c}}

\newcommand{\ft}{F_{2}}
\newcommand{\fl}{F_{\mathrm{L}}}
\newcommand{\nf}{n_\mathrm{f}}
\newcommand{\seq}{\sum_{q}^{\nf} e_q^2}

\newcommand{\cf}{C_\mathrm{F}}
\newcommand{\TR}{T_\mathrm{R}}
\newcommand{\seqav}{ \bar{e}_q^2}
\newcommand{\cgf }{ C}

\newcommand{\cflg}{ {C}_{\fl g}}
\newcommand{\cfls }{ {C}_{\fl \Sigma}}
\newcommand{\wft}{\widetilde{F}_2}
\newcommand{\cflwft}{ {C}_{\fl \wft}}
\newcommand{\wftp}{\widetilde{F'}_2}
\newcommand{\wfl}{\widetilde{F}_{\mathrm{L}}}
\newcommand{\wflp}{\widetilde{F'}_{\mathrm{L}}}
\newcommand{\wflpp}{\widetilde{F''}_{\mathrm{L}}}
\newcommand{\ftw }{F_2^{\rm W^-}}
\newcommand{\wftw}{\widetilde{F}_2^{\rm W^-}}
\newcommand{\fk }{F_3}
\newcommand{\fkw }{F_3^{\rm W^-}}
\newcommand{\ftcw }{F_{2\rm c}^{\rm W^-}}
\newcommand{\wftcw}{\widetilde{F}_{2\rm c}^{\rm W^-}}
\newcommand{\xq }{ (x, Q^2)}
\newcommand{\eu }{ e_u^2}
\newcommand{\ed }{ e_d^2}
\newcommand{\es }{ e_s^2}

\usepackage[breaklinks,colorlinks,citecolor=citcolor,urlcolor=blue,linkcolor=lcolor]{hyperref}

\definecolor{lcolor}{rgb}{0.5,0,0}
\definecolor{citcolor}{rgb}{0,0.3,0.0}

\newcommand{\as}{\alpha_\mathrm{s}}

\begin{document}

\title{Evolution of structure functions in momentum space}

\author{Tuomas Lappi}
\email{tuomas.v.v.lappi@jyu.fi}
\author{Heikki Mäntysaari}
\email{heikki.mantysaari@jyu.fi}
\author{Hannu Paukkunen}
\email{hannu.paukkunen@jyu.fi}
\author{Mirja Tevio}
\email{mirja.h.tevio@jyu.fi}
\affiliation{
University of Jyväskylä, Department of Physics,  P.O. Box 35, FI-40014 University of Jyväskylä, Finland
}
  
\affiliation{
Helsinki Institute of Physics, P.O. Box 64, FI-00014 University of Helsinki, Finland
}

\begin{abstract}

We formulate the momentum-space Dokshitzer-Gribov-Lipatov-Altarelli-Parisi (DGLAP) evolution equations for structure functions measurable in deeply inelastic scattering. We construct a six-dimensional basis of structure functions that allows for a full three flavor structure and thereby provides a way to calculate perturbative predictions for physical cross sections directly without unobservable parton distribution functions (PDFs) and without the associated scheme dependence. We derive the DGLAP equations to first non-zero order in strong coupling $\as$, but the approach can be pursued to arbitrary order in perturbation theory. We also numerically check our equations against the conventional PDF formulation.  

\end{abstract}

\maketitle

\section{Introduction}
\label{sec:intro}

The factorization of short- and long-range physics \cite{Collins:1989gx} in hard-process cross sections constitutes the cornerstone of contemporary collider physics. While the short-range parts can be systematically calculated by the perturbative techniques of Quantum Chromodynamics (QCD) and electroweak theory, the long-range physics is related to the nonperturbative structure of QCD bound states, and ultimately requires input from experimental measurements. The typical way to describe this long range physics is in terms of parton distribution functions (PDFs), which can, at least at leading order, be given an intuitive physical interpretation as parton densities. 

The typical way to proceed is to start from a perturbative calculation of a cross section for a specific process. The perturbative cross section is then split into short- and long-distance contributions. The latter are factorized into the PDFs, which then become dependent on a factorization scale $\mu$. The dependence on $\mu$ shows up as logarithmic terms in the coefficient functions. 
The dependence of the PDFs on $\mu$ leads to  the Dokshitzer-Gribov-Lipatov-Altarelli-Parisi (DGLAP)~\cite{Gribov:1972ri,Lipatov:1974qm,Altarelli:1977zs,Dokshitzer:1977sg} renormalization group equations for the PDFs. Solving the DGLAP equation, followed by setting the value of the factorization scale to the relevant physical momentum scale such that the logarithmic terms in the coefficient functions remain small, performs a resummation of large logarithmic corrections to the cross section.  After this is done, one  usually extracts the initial conditions for the DGLAP evolution of the PDFs from global fits of experimental data~\cite{Ethier:2020way}.

The split of the cross sections into short- and long-distance parts is, however, not unique at any given order in perturbation theory. In addition to the dependence on the factorization and renormalization scales, it depends on the choice of the factorization and renormalization schemes. Formally, this dependence is always of higher order in the QCD coupling $\as$ than the order of $\as$ to which the cross section was calculated, but it can still be numerically non-negligible. It is customary to consider the sensitivity of the cross sections to variations of the factorization scale $\mu$ and the coupling renormalization scale $\mu_r$ as an estimate for the theoretical uncertainty due to missing higher-order contributions. Although the change from one factorization scheme to another can also be calculated perturbatively, its effect is rarely quantified in practice. The standard choice is nowadays the so-called $\overline{\rm MS}$ scheme~\cite{Bardeen:1978yd} and all the publicly available PDFs and codes to calculate higher-order cross sections tend to adopt this particular choice. As a result, the sensitivity of the cross sections to the choice of scheme is hardly ever considered. 

In addition to the choice of scheme, there is considerable freedom in parametrizing the initial conditions for the DGLAP evolution of PDFs. Since the PDFs themselves are not physical observables, different functional forms can in fact lead to similar values for physical observables. Thus PDF sets can often have larger errors than the experimental data that they have been determined from. The PDFs resulting from fits to data at different orders in $\as$ or different factorization schemes can look very different and are difficult to compare to each other in any meaningful way. Since PDFs are not physical observables, it is not very clear to what extent one should e.g. favor PDFs that are smooth functions, and there is a continuing discussion on  whether  PDFs should be positive-definite~\cite{Altarelli:1998gn,Candido:2020yat,Collins:2021vke, Candido:2023ujx}.
In the end, perturbative QCD and DGLAP evolution make specific predictions about the dependence of measurable cross sections on kinematical variables. When phrased in terms of PDFs the comparison between calculations and measurements is less direct~\cite{Armesto:2022mxy}, and one might prefer to have a way to calculate the effect of DGLAP evolution directly in terms of cross sections.

We advocate in this paper in favor of  an alternative to  the conventional PDF-based approach, and formulate factorization directly in terms of observable quantities. A simple choice, and the one we will adopt here, is to use structure functions in deeply inelastic scattering (DIS). In the usual approach the DIS structure functions $F_i(x,Q^2)$ are expressed in terms of PDFs $f_j(\mu^2)$ schematically as
\begin{equation}
F_i(x,Q^2) = \sum_j C_{ij}(Q^2,\mu^2) \otimes f_j(\mu^2) \,, \label{eq:Fitf}
\end{equation}
where $C_{ij}$ denote the scheme-dependent coefficient functions, $x$ is the Bjorken variable, and $Q^2$ the squared momentum transfer in the process. However, one can -- and this we will explicitly demonstrate in the present manuscript for a certain choice of structure functions -- also express the PDFs in terms of structure functions,
\begin{equation}
f_i(x,\mu^2) = \sum_j C^{-1}_{ij}(Q^2,\mu^2) \otimes F_j(Q^2) \,, \label{eq:fitF}
\end{equation}
where $C^{-1}_{ij}$ denote the perturbative inverse of $C_{ij}$. Taking the $Q^2$ derivative of Eq.~(\ref{eq:Fitf}) and using Eq.~(\ref{eq:fitF}) then leads to a set of DGLAP-type evolution equations,  
\begin{equation}
\frac{\dd{F_i(x,Q^2)}}{\dd{\log Q^2}} = \sum_j P_{ij} \otimes F_j(Q^2) \,, \label{eq:Fe}
\end{equation}
which predict the behaviour of the physical structure functions as a function of the kinematic variable $Q^2$. 
The main advantage of Eq.~(\ref{eq:Fe}) is that the evolution kernels $P_{ij}$ cannot depend on the factorization scheme nor on the factorization scale \cite{vanNeerven:1999ca}: Being physical observables, the structure functions $F_i$ are independent of these choices, and Eq.~(\ref{eq:Fe}) then implies that the evolution kernels must be so as well. For example, there cannot be any $\mu$-dependent terms, since there are no PDFs that would compensate for the non-zero derivative. As the evolution kernels are power series in $\as$, at any finite perturbative order they still depend on the renormalization scale and scheme, though the dependence is of higher order in $\as$ than to which the evolution kernels were calculated.

In addition, the initial condition for the evolution in Eq.~(\ref{eq:Fe}) could -- in the ideal case -- be directly given by experimental data at fixed $Q^2$ and this condition would remain the same irrespective of the perturbative order to which the evolution kernels are calculated. This in contrast to the conventional approach in which the PDFs at leading order (LO), next-to-LO (NLO) and next-to-NLO (NNLO) can be mutually very different. In practice, all necessary structure functions are not (and never will be) available at constant $Q^2$ and thus the non-perturbative input must still be parametrized and fitted. However, even then the mere fact that one works with obervable cross sections means that one  neatly evades e.g. thorny questions such as 
the one concerning the positivity of the PDFs \cite{Altarelli:1998gn,Candido:2020yat,Collins:2021vke, Candido:2023ujx}. The possible evidence for the small-$x$ resummation \cite{Ball:2017otu,xFitterDevelopersTeam:2018hym} could also be put to a much stronger footing in the physical-basis approach as there is no room to mimic the genuine small-$x$ effects through setting a $x$-dependent factorization scale as e.g. done in Ref.~\cite{Hou:2019efy}. 

In addition, it is reasonable to assume that physical structure functions behave smoothly, whereas the PDFs could, in principle, have discontinuities and be even fractals. Since the PDFs can be expressed in terms of physical structure functions it follows that also all other processes e.g. at the LHC can be directly expressed in terms of DIS structure functions, simply substituting Eq.~(\ref{eq:fitF}) into the expressions of cross sections.  For example, one might consider the cross section of Higgs production by gluon fusion~\cite{LHCHiggsCrossSectionWorkingGroup:2013rie}
\begin{equation}
    \label{eq:HiggsproductionPDFbasis}
    \sigma(p+p\longrightarrow {\rm H}+X) = \int\dd x_1\dd x_2 g(x_1,\mu)g(x_2,\mu)\hat \sigma_{gg \rightarrow H + X}(x_1, x_2,\frac{m_H^2}{\mu^2}),
\end{equation}
where $m_H$ is the Higgs mass, $g(x_1,\mu)$ and $g(x_2,\mu)$ are the gluon PDFs of the proton, and $\hat \sigma_{gg \rightarrow H + X}(x_1, x_2,  m_H^2 / \mu^2)$ denotes the collinear-renormalized partonic cross section. 
In terms of structure functions, 
\begin{equation}
    \label{eq:HiggsproductionPhysBasis}
    \sigma(p+p\longrightarrow H+j) = \int\dd x_1\dd x_2 \hat \sigma_{gg \rightarrow H + X}(x_1, x_2,\frac{m_H^2}{\mu^2})
    \left[\sum_j C^{-1}_{gj}(Q^2,\mu^2) \otimes F_j(Q^2)\right]_{x_1} 
    \left[\sum_k C^{-1}_{gk}(Q^2,\mu^2) \otimes F_k(Q^2)\right]_{x_2}  ,
\end{equation}
where 
the subscripts in $[\cdots]_{x}$ indicate the $x$ values in the convolution. 
Here the factorization scale dependence in the partonic cross section $\hat \sigma_{gg \rightarrow H + X}(x_1, x_2, m_H^2 / \mu^2)$ appears only as logarithmic fractions $\log\left(m^2_H/\mu^2\right)$. Similarly, the factorization scale dependence in the inverse coefficient functions $C^{-1}_{gj}(Q^2,\mu^2)$ appears in terms $\log \left(Q^2/\mu^2\right)$.  
It is shown in Ref.~\cite{Harland-Lang:2018bxd} that   
the explicit $\mu$ dependence vanishes and terms 
$\log\left(Q^2/m_H^2\right)$ 
are left behind. Indeed this is required by the fact that the the physical cross section $\sigma(p+p\longrightarrow H+j)$ cannot depend on an arbitrary factorization scale.
 This means that in  Eq.~\eqref{eq:HiggsproductionPhysBasis} one does not have to specify an exact relation between the $\mu^2$ and the physical scales $Q^2$ and $m^2_H$. This is an advantage of the physical basis, compared to the usual PDF-based approach, where one has to choose the value of the factorization scale, which typically is set as $Q^2 = \mu^2$. 

The notion of a physical basis is already quite an old one: it was first discussed in the 1980's \cite{Furmanski:1981cw} and has thereafter resurfaced in different contexts \cite{Catani:1996sc,Blumlein:2000wh,Hentschinski:2013zaa,Harland-Lang:2018bxd,Blumlein:2021lmf,vanNeerven:1999ca,RuizArriola:1998er}
(see also discussion in Refs.~\cite{Baulieu:1979mr} and \cite{Floratos:1980hm}). 
Whereas the existing literature on the subject is restricted to special cases or case studies, the novelty of the present paper is that we present a full dimension-six physical basis of structure functions which corresponds to PDFs with three active parton flavors. In principle, this constitutes a basis that would allow for a global analysis of world DIS and collider data. We present the evolution kernels directly in momentum space -- though at this stage only at the first non-zero order in $\as$. 

The structure of the paper is the following. In Sec.~\ref{sec:DGLAPevolutioninphysicalbasis} we outline the underlying idea of the physical basis in a simplified case by first working out an example with no flavor separation. In particular, we show how the inversion of the coefficient functions, as in Eq.~\eqref{eq:fitF}, works directly in momentum $x$, as opposed to Mellin space. In Sec.~\ref{sec:DGLAPevolutioninphysicalbasiswithvalencequarks} we present the evolution equations that include the flavor separation as well. Finally, we draw our conclusions in Sec.~\ref{sec:conclusions} outlining the steps towards higher-order corrections and a global analysis.

\section{Evolution of a two-observable physical basis}
\label{sec:DGLAPevolutioninphysicalbasis}

To highlight the underlying idea of the physical-basis approach, let us first consider a toy model of two independent observables. We choose these two to be the DIS structure functions $\ft$  and $\fl$ accounting only for the massless quark singlet~\cite{Blumlein:2012bf}, 
\begin{equation}
    \label{eq:singlet}
    \Sigma(x, \mu_f^2) = \sum_{q}\left[q(x, \mu_f^2)+\overline{q}(x, \mu_f^2)\right],
\end{equation} 
and the gluon PDF $g(x,\mu_f^2)$. We do not yet take a specific value for the is the factorization scale  $\mu_f$ at this point. In this approximation, we can write the structure functions $\ft$  and $\fl$ up to the first two non-trivial orders in $\as$ in terms of PDFs as, 
\begin{align}
\frac{1}{\seqav} \frac{F_2(x, Q^2)}{x} & = \bigg\{C_{F_2\Sigma}^{(0)} + \frac{\as(\mu^2_r)}{2\pi} \bigg[ C_{F_2\Sigma}^{(1)} - \log\left(\frac{{\mu_f^2}}{{Q^2}}\right) C_{F_2\Sigma}^{(0)} \otimes P_{qq} \bigg] \bigg\} \otimes \Sigma(x, \mu_f^2) \label{eq:F2master} \\
& \hspace{1.2cm} + 2n_f \frac{\as(\mu^2_r)}{2\pi} \bigg[C_{F_2g}^{(1)} - \log\left(\frac{{\mu_f^2}}{{Q^2}}\right) C_{F_2\Sigma}^{(0)} \otimes P_{qg} \bigg] \otimes g(x, \mu_f^2) \,, \nonumber \\
\frac{1}{\seqav} \frac{F_{\rm L}(x, Q^2)}{x} & = \frac{\as(\mu^2_r)}{2\pi} \bigg\{C_{F_{\rm L}\Sigma}^{(1)} + \frac{\as(\mu^2_r)}{2\pi} \bigg[ C_{F_{\rm L}\Sigma}^{(2)} - \log\left(\frac{{\mu_f^2}}{{Q^2}}\right) C_{F_{\rm L}\Sigma}^{(1)} \otimes P_{qq} 
- 2n_f \log\left(\frac{{\mu_f^2}}{{Q^2}}\right) C_{F_{\rm L}g}^{(1)} \otimes P_{\rm gq}
\bigg] \bigg\} \otimes \Sigma(x, \mu_f^2) \label{eq:FLmaster} \\
& \hspace{0.2cm} + 2n_f \frac{\as(\mu^2_r)}{2\pi} \bigg\{ C_{F_{\rm L}g}^{(1)} + \frac{\as(\mu^2_r)}{2\pi} \bigg[ C_{F_{\rm L}g}^{(2)} - \log\left(\frac{{\mu_f^2}}{{Q^2}}\right) C_{F_{\rm L}\Sigma}^{(1)} \otimes P_{qg} 
- \log\left(\frac{{\mu_f^2}}{{Q^2}}\right) C_{F_{\rm L}g}^{(1)} \otimes P_{\rm gg}
\bigg] \bigg\} \otimes g(x, \mu_f^2) \nonumber \\
& \hspace{0.1cm} + \left(\frac{\as(\mu^2_r)}{2\pi}\right)^2 \left[b_0 \log\left(\frac{{\mu_r^2}}{{Q^2}}\right) \right] 
\bigg[C_{F_{\rm L}\Sigma}^{(1)} \otimes \Sigma(x, \mu_f^2) + 2n_f C_{F_{\rm L}g}^{(1)} \otimes g(x, \mu_f^2) \bigg] \,. \nonumber 
\end{align}
In these expressions $\mu_r$ denotes the renormalization scale, $n_f$ is the number of massless flavours, $b_0$ is the first coefficient in the QCD $\beta$ function, 
\begin{equation}
b_0 = \frac{11C_A-4T_Rn_f}{6} \,, \ C_A = 3 \,, \ T_R = 1/2 \,,
\end{equation}
and $\seqav$ is the average quark charge,  
\begin{equation}
    \label{eq:eq2av}
   \seqav\equiv \frac{1}{\nf}\sum_q e_q^2 \,, 
\end{equation}
where $e_q$ denotes the electric charge of quark $q$. The LO splitting functions are given by 
\begin{align}
\label{eq:Pqq}
P_{qq}(z) & = \cf\left[\frac{1+z^2}{(1-z)_{+}}+\frac{3}{2}\delta(1-z) \right] \,, \\ 
P_{qg}(z) & = \TR \left[z^2+(1-z)^2\right]  \,, \\
P_{gg}(z) & =  2\nc \left[\frac{1-z}{z}+z(1-z)+\frac{z}{(1-z)_{+}}\right] + b_0\delta(1-z)  \,, \\
\label{eq:Pgq}
P_{gq}(z) & = \cf\frac{1+(1-z)^2}{z} \,,
\end{align}
where $N_c=C_A=3$, $\cf=(\nc^2-1)/(2\nc)$, and the plus function is defined as 
\begin{align}
    \label{eq:plusf}
    \int_x^1 \dd z \frac{f(z)}{(1-z)_+} & = \int_x^1\dd z \frac{f(z)-f(1)}{1-z}+f(1)\log(1-x).
\end{align}
We work in this paper at the first non-zero order in $\as$, meaning that for each physical observable we take the first term of the coefficient function, $\ft\sim \as^0$ and $\fl \sim \as^1$.  
For the two-observable basis these first non-zero coefficient functions are
\begin{align}
C_{F_2\Sigma}^{(0)}(z) & =  \delta(1-z) \,, \\ 
\cfls^{(1)}(z) & =  2 \cf z \,, \label{eq:C1FLS} \\ 
\cflg^{(1)}(z) & =  4 \TR z\left(1-z\right) \label{eq:C1FLg} \,. 
\end{align}
The symbol $\otimes$ denotes the usual convolution 
\begin{equation}
\label{eq:convolution2}
f\otimes g \equiv \int_{x}^{1} \frac{\dd z}{z}f(z)g\left(\frac{x}{z}\right) \,. 
\end{equation}

At this point one would typically differentiate Eqs.~(\ref{eq:F2master}) and~(\ref{eq:FLmaster}) with respect to the scale $\mu_f^2$ to derive the DGLAP equation satisfied by the PDFs. One would then set $\mu_f^2$ to express the physical the structure functions in terms of the evolved PDFs. Here our emphasis is slightly different: we want to derive an evolution equation for the $Q^2$ dependence of the structure functions by differentiating with respect to $Q^2$. Before doing that we must  invert the leading non-zero order part of the relations  Eq.~(\ref{eq:F2master}) and Eq.~(\ref{eq:FLmaster}) and  express the quark singlet and gluon PDFs in terms of the structure functions $F_2$ and $F_{\rm L}$. The result of this exercise is
\begin{align}
    \Sigma(x,\mu_f^2) & = \frac{1}{\seqav} 
    \wft\xq
        \label{eq:sigmafromF2} \,, \\ 
    g(x, \mu_f^2) & = \frac{1}{\nf\seqav} \bigg( \cgf_{g\wftp}\otimes \wftp +\cgf_{g\wft}\otimes \wft +\cgf_{g\wflpp}\otimes\wflpp+\cgf_{g\wflp}\otimes\wflp+\cgf_{g\wfl}\otimes\wfl  \bigg) \,, \label{eq:gfromF2FL} 
\end{align}
where
\begin{align}
\wft\xq & \equiv \frac{\ft(x, Q^2)}{x} \,, \\
\label{eq:wfl}
\wfl\xq &\equiv  \frac{2\pi}{\as(\mu_r^2)} \frac{\fl(x, Q^2)}{x} \,, \\
\label{eq:f2lprime}
\widetilde{F'}_{2,L}\xq &\equiv  x\frac{\dd}{\dd{x}}\widetilde{F}_{2,L}(x, Q^2) \,, \\
\label{eq:wflpp}
\wflpp\xq &\equiv x^2\frac{\dd[2]}{\dd{x^2}}\wfl(x, Q^2) \,,
\end{align}
and 
\begin{align}
\label{eq:Cgwftp}
\cgf_{g\wftp}(z) &\equiv \frac{\cf}{4\TR} \delta(1-z) \,, \\ 
\label{eq:Cgwft}
\cgf_{g\wft}(z) &\equiv -\frac{\cf}{2\TR}\delta(1-z) \,, \\  
\label{eq:Cgwflpp}
\cgf_{g\wflpp}(z) &\equiv \frac{1}{8\TR}\delta(1-z) \,, \\
\label{eq:Cgwflp}
\cgf_{g\wflp}(z) & \equiv -\frac{1}{4\TR}\delta(1-z) \,, \\  \label{eq:Cgwfl}
\cgf_{g\wfl}(z) & \equiv \frac{1}{4\TR}\delta(1-z)  \,. 
\end{align}

The evolution equations now follow by taking the $Q^2$ derivative in Eqs.~(\ref{eq:F2master}) and (\ref{eq:FLmaster}), and then using Eqs.~(\ref{eq:sigmafromF2}) and (\ref{eq:gfromF2FL}), 
\begin{align}
\label{eq:toyf2eq}
 \frac{\dd}{\dd\log Q^2} \Bigg[ \frac{F_2(x, Q^2)}{x} \bigg] & = 
      \frac{\as(\mu^2_r)}{2\pi} \bigg[ C_{F_2\Sigma}^{(0)} \otimes P_{qq}  \otimes 
    \wft
            \\
 & + 2  C_{F_2\Sigma}^{(0)} \otimes P_{qg} \otimes \bigg( \cgf_{g\wftp}\otimes \wftp +\cgf_{g\wft}\otimes \wft +\cgf_{g\wflpp}\otimes\wflpp+\cgf_{g\wflp}\otimes\wflp+\cgf_{g\wfl}\otimes\wfl  \bigg) \Bigg] \nonumber \,, \\
 \frac{\dd}{\dd\log Q^2} \bigg[ \frac{F_{\rm L}(x, Q^2)}{x} \bigg] & = 
 \left(\frac{\as(\mu^2_r)}{2\pi}\right)^2 \bigg[   C_{F_{\rm L}\Sigma}^{(1)} \otimes P_{qq} 
+  2n_f C_{F_{\rm L}g}^{(1)} \otimes P_{\rm gq} - b_0 C_{F_{\rm L}\Sigma}^{(1)}
 \bigg] \otimes 
 \wft 
 \nonumber \\
& + 2 \left(\frac{\as(\mu^2_r)}{2\pi}\right)^2 \bigg[  C_{F_{\rm L}\Sigma}^{(1)} \otimes P_{qg} 
+ C_{F_{\rm L}g}^{(1)} \otimes P_{\rm gg} -b_0 C_{F_{\rm L}g}^{(1)}
 \bigg] \\
 & \otimes \bigg( \cgf_{g\wftp}\otimes \wftp +\cgf_{g\wft}\otimes \wft +\cgf_{g\wflpp}\otimes\wflpp+\cgf_{g\wflp}\otimes\wflp+\cgf_{g\wfl}\otimes\wfl  \bigg) \nonumber \,. 
\end{align}
By setting the renormalization scale equal to the momentum transfer, $\mu_r^2 = Q^2$, and using the LO renormalization group equation,

\begin{equation}
\mu^2_r \frac{\dd\as(\mu^2_r)}{\dd\mu^2_r} = - \left(\frac{b_0}{2\pi}\right) \alpha^2_s(\mu^2_r) \,, 
\end{equation}
the equation for $F_{\rm L}$ can be written as, 
\begin{align}
 \frac{\dd}{\dd\log Q^2} \bigg[ \frac{2\pi}{\as(Q^2)} \frac{F_{\rm L}(x, Q^2)}{x} \bigg] & = 
 \left(\frac{\as(Q^2)}{2\pi}\right) \bigg[   C_{F_{\rm L}\Sigma}^{(1)} \otimes P_{qq} 
+  2n_f C_{F_{\rm L}g}^{(1)} \otimes P_{\rm gq} 
 \bigg] \otimes 
 \wft 
 \nonumber \\
& + 2 \left(\frac{\as(Q^2)}{2\pi}\right) \bigg[  C_{F_{\rm L}\Sigma}^{(1)} \otimes P_{qg} 
+ C_{F_{\rm L}g}^{(1)} \otimes P_{\rm gg}  \bigg] \label{eq:toyfleq} \\
 & \otimes \bigg( \cgf_{g\wftp}\otimes \wftp +\cgf_{g\wft}\otimes \wft +\cgf_{g\wflpp}\otimes\wflpp+\cgf_{g\wflp}\otimes\wflp+\cgf_{g\wfl}\otimes\wfl  \bigg) \nonumber \,.
\end{align}
These forms of the equations demonstrate the general idea of the inversion procedure. However, 
they now contain up to three convolution integrals. Since the coefficient and splitting functions appearing in them are known analytically, these can be reduced to one by analytically performing part of the convolution integrals. The final, explicit forms of the evolution equations for $F_{2}$ and $F_{\rm L}$ read

\begin{equation}
    \label{eq:F2dglap}
      \frac{\dd \ft(x, Q^2)}{\dd \log Q^2} = \frac{\as(Q^2)}{2\pi}x\Big\{P^{2\text{-}o.}_{\ft \wft}\otimes  \wft +P^{2\text{-}o.}_{\ft\wfl}\otimes\wfl+P^{2\text{-}o.}_{\ft\wflp}\otimes \wflp\Big\} \,, 
   \end{equation}

\begin{equation}
    \label{eq:FLdglap}
     \frac{\dd}{\dd \log Q^2}\left(\frac{ \fl(x, Q^2)}{\frac{\as(Q^2)}{2\pi}}\right) = \frac{\as(Q^2)}{2\pi}x\Big\{P^{2\text{-}o.}_{\fl \wft}\otimes  \wft +P^{2\text{-}o.}_{\fl\wfl}\otimes\wfl+P^{2\text{-}o.}_{\fl\wflp}\otimes \wflp\Big\} \,, 
   \end{equation}
where the  kernels are defined by
\begin{align}
    \label{eq:PF2F2}
    P^{2\text{-}o.}_{\ft \wft}(z) & \equiv \cf \left[\delta(1-z) -(1-z)+\frac{1+z^2}{(1-z)_+}\right]\,,\\
    \label{eq:PF2FL}
    P^{2\text{-}o.}_{\ft\wfl}(z) &\equiv \frac{1}{4}\delta(1-z)+\frac{1}{2}\,, \\
    \label{eq:PF2dFL}
    P^{2\text{-}o.}_{\ft\wflp}(z) & \equiv -\frac{1}{4}\delta(1-z)\,,\\
    \label{eq:PFLF2}
    P^{2\text{-}o.}_{\fl \wft}(z) & \equiv \cf \Bigg[4\nc-8\TR\nf+2\left(\frac{4\TR\nf}{3}-\nc\right)\frac{1}{x}+2\left(\cf+\nc-b_0 \right)z+4\left(\frac{4\TR\nf}{3}-\nc\right)z^2 \nonumber\\
    &+4\left(2\nc -\cf-2\TR\nf \right)z\log\left(z\right)+4\left(\cf-\nc\right)z\log\left(1-z\right) \Bigg] \,,\\
    \label{eq:PFLFL}
    P^{2\text{-}o.}_{\fl\wfl}(z) &\equiv  \left[\frac{\cf}{2} +b_0 \right]\delta(1-z)  +\Big[\cf-2\nc+2\nc\frac{1}{z}+(4\nc-\cf)z+4\nc z^2\log\left(1-z\right)\Big] \,, \\
    \label{eq:PFLdFL}
    P^{2\text{-}o.}_{\fl\wflp}(z) & \equiv -2\nc z^2\log\left(1-z\right)\,.
\end{align}
Here the superscript $2\text{-}o.$ refers to the two observable physical basis in which we are working. The derivatives appearing in Eqs.~\eqref{eq:toyf2eq} and~\eqref{eq:toyfleq} have been removed by partial integrations. However, after partially integrating the second derivatives, one first derivative still remains.

\begin{figure}[ht]
  \includegraphics[width=.5\linewidth]{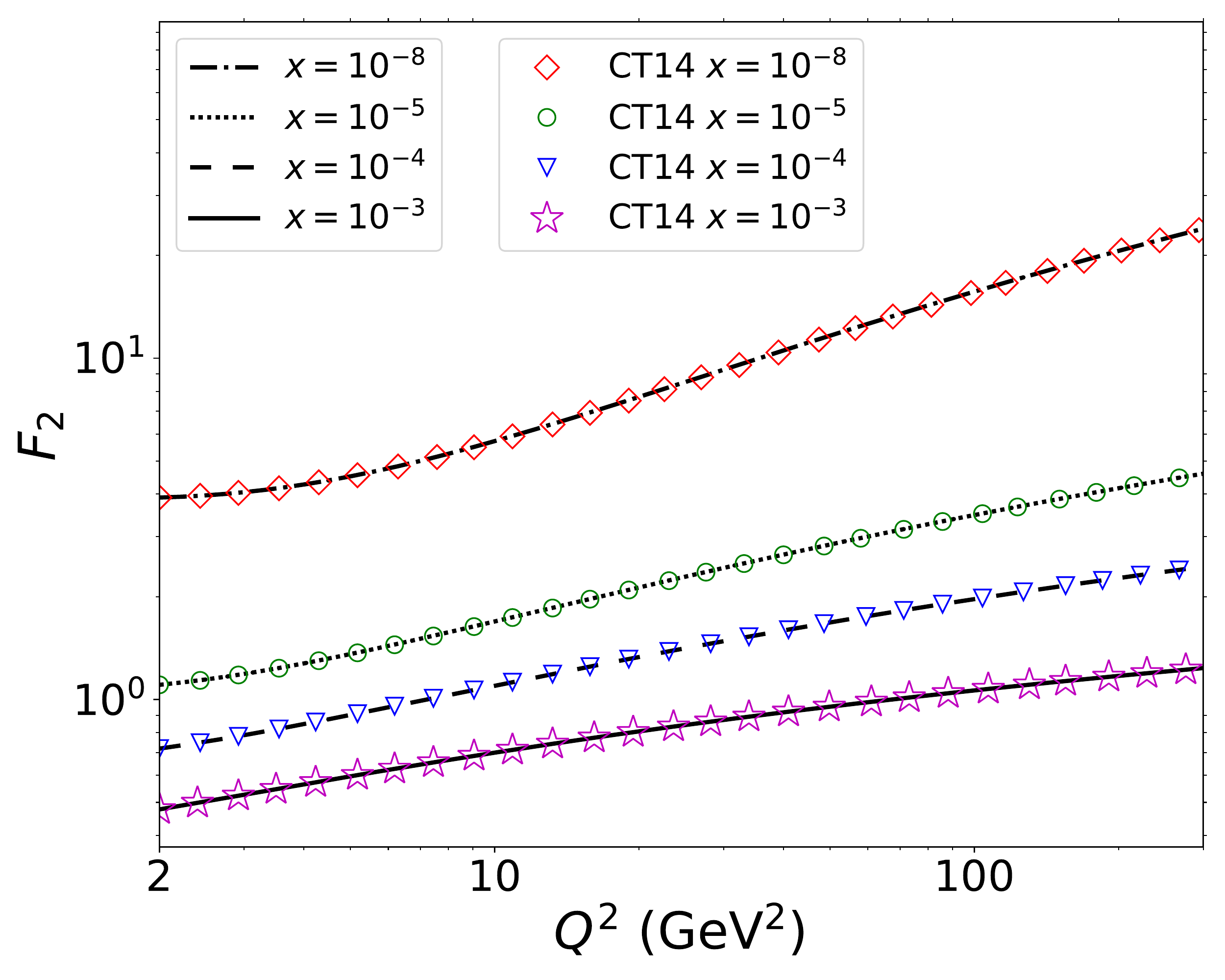}\hfill
  \includegraphics[width=.5\linewidth]{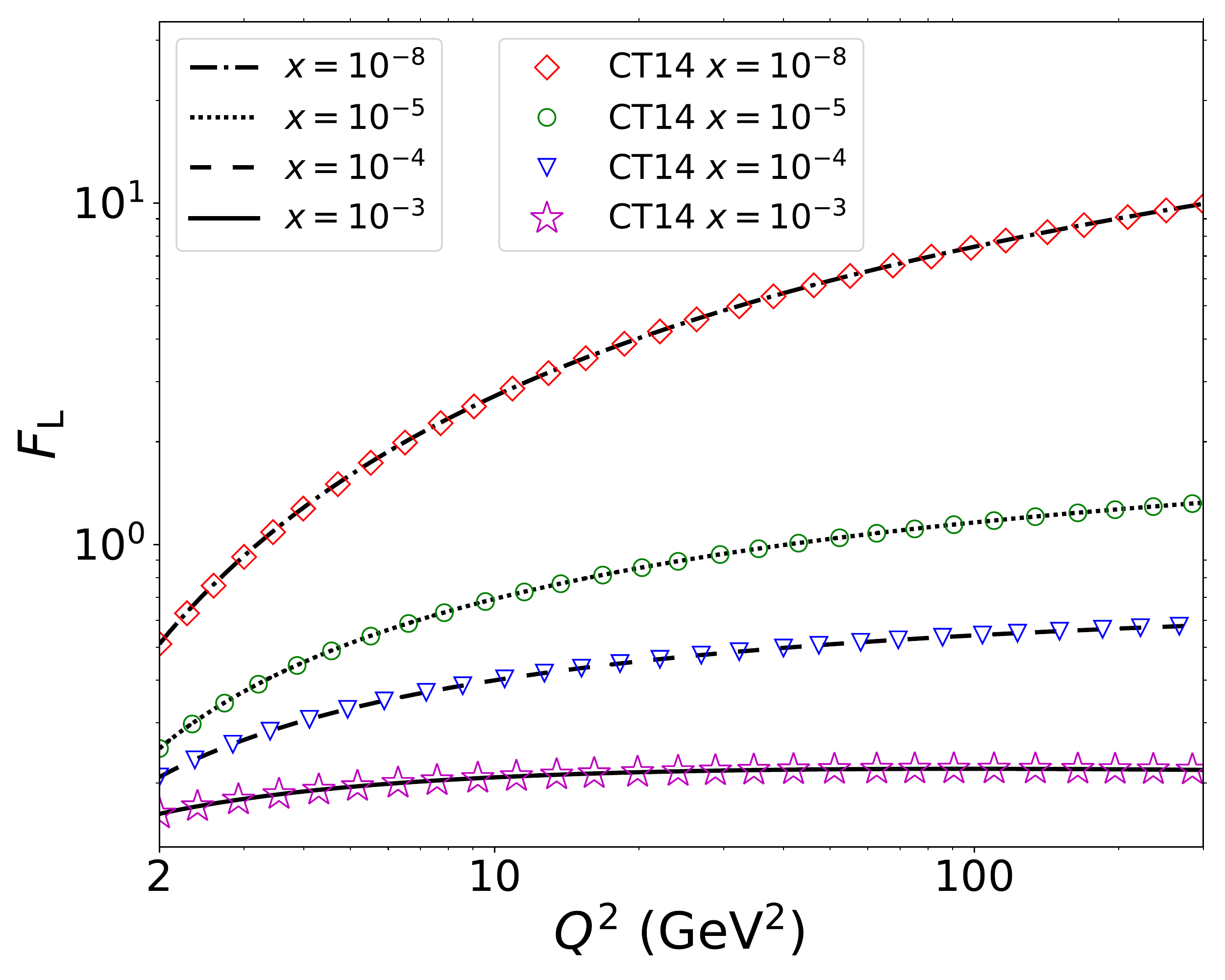}
  \includegraphics[width=.5\linewidth]{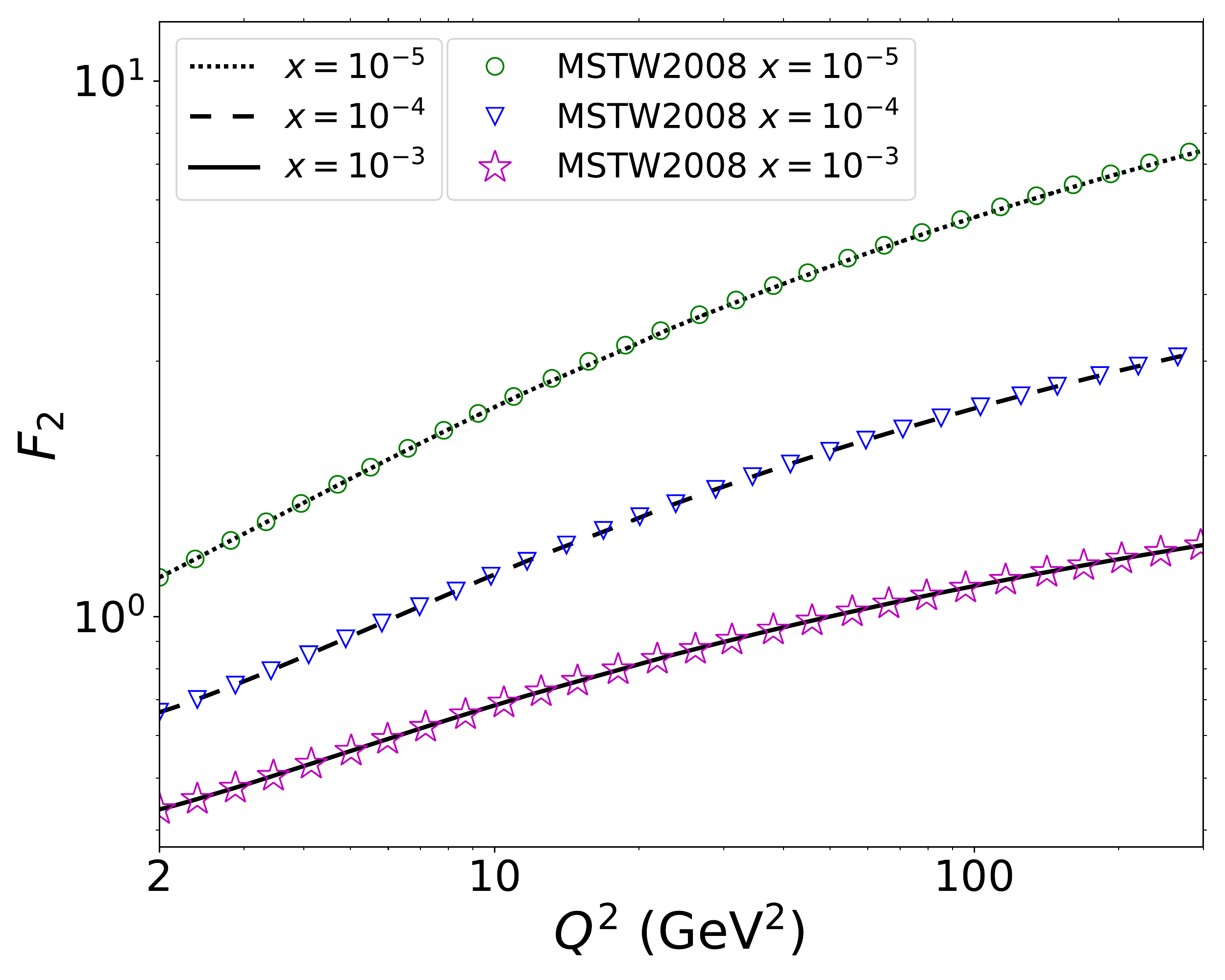}\hfill
  \includegraphics[width=.5\linewidth]{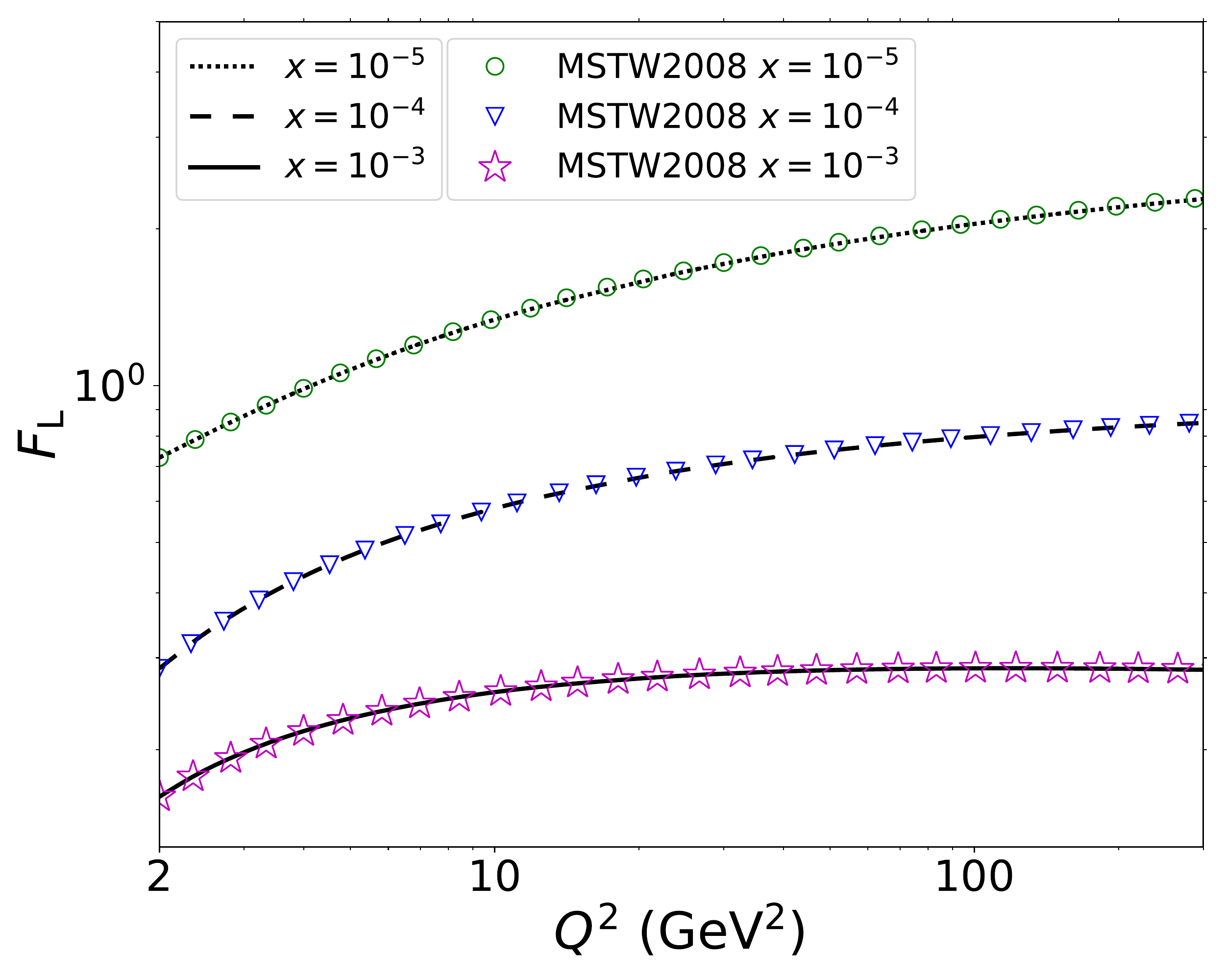}
  \includegraphics[width=.5\linewidth]{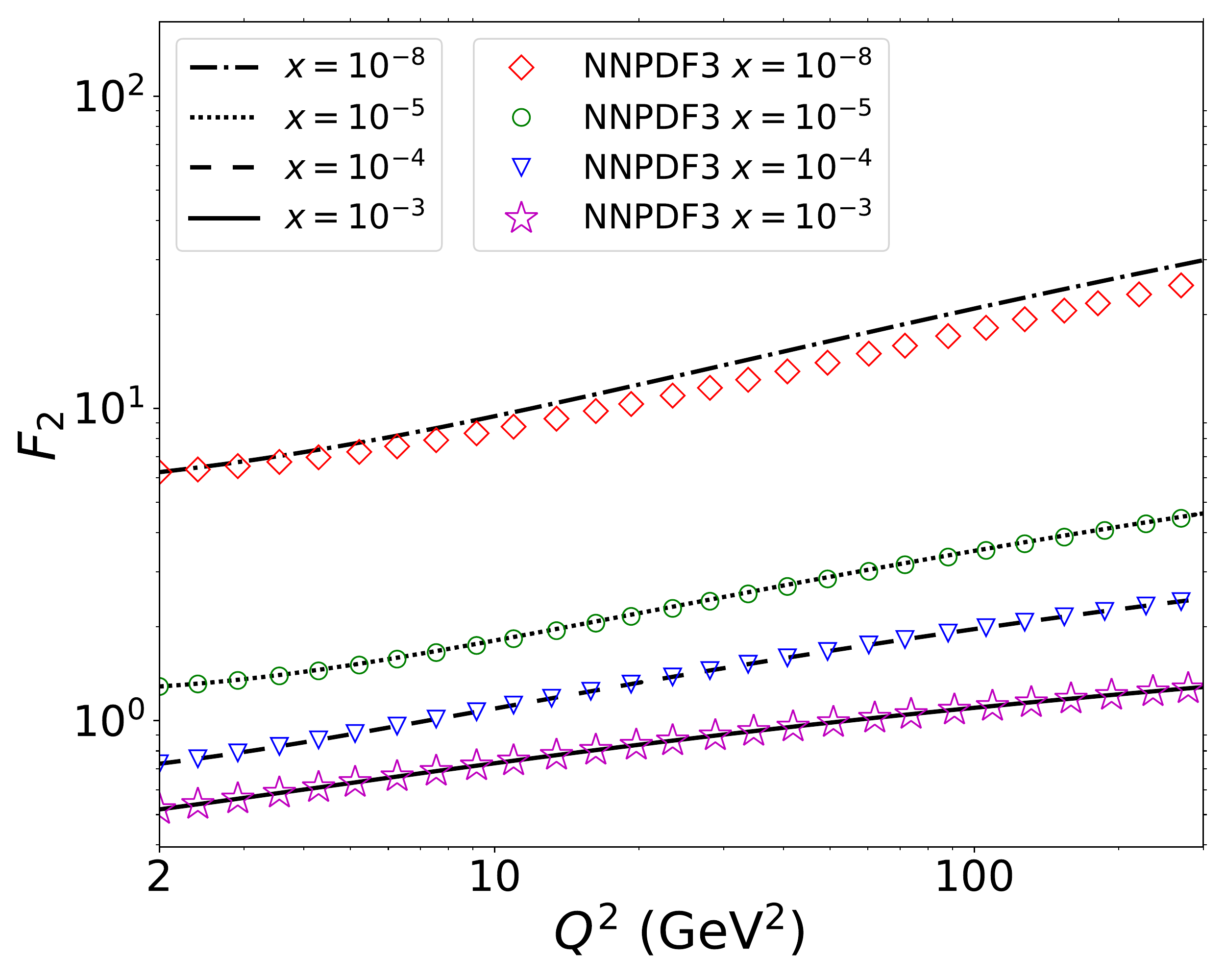}\hfill
  \includegraphics[width=.5\linewidth]{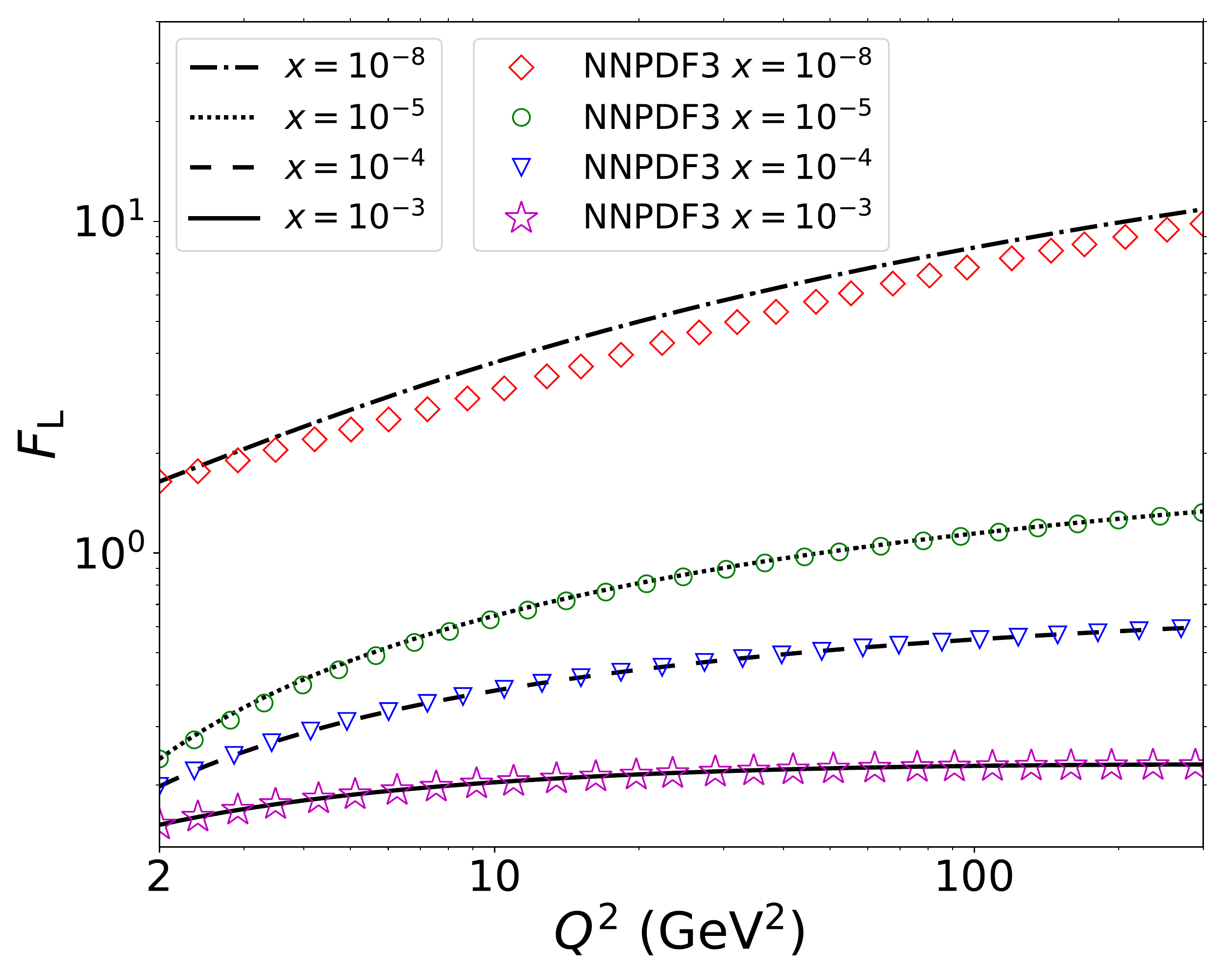}
  \caption{The $Q^2$ dependence of $\ft$ (left) and $\fl$ (right) using the physical-basis approach (curves) compared with the usual PDF-based approach (markers). The comparisons in the upper-most panels correspond to the CT14lo\_NF3 PDFs, in the middle panels to the MSTW2008lo68cl\_nf3 PDFs, and in the lower-most panels to  the NNPDF30\_lo\_as\_0118\_nf\_3 PDFs.}
\label{fig:results}
\end{figure}

We note that we could also arrive with Eqs.~(\ref{eq:F2dglap}) and (\ref{eq:FLdglap}) by setting $\mu^2_r=\mu^2_f=Q^2$ already in Eqs.~(\ref{eq:F2master}) and (\ref{eq:FLmaster}), taking the $\log Q^2$ derivative, and using the LO DGLAP equations for PDFs, 
\begin{align}
    \label{eq:quarkdglap}
    \frac{\dd q(x, Q^2)}{\dd\log Q^2} & = \frac{\as(Q^2)}{2\pi}\Big(P_{qq}\otimes q+ P_{qg}\otimes g \Big), \\
    \label{eq:gdglap}
    \frac{\dd{g(x, Q^2)}}{\dd \log Q^2} & = \frac{\as(Q^2)}{2\pi}\Big(P_{gq}\otimes\Sigma+P_{gg}\otimes g \Big). 
\end{align}
However, the derivation presented above highlights the fact that it is not in fact necessary to choose a specific value for the factorization scale. In fact,  Eqs.~(\ref{eq:F2dglap}) and (\ref{eq:FLdglap}) really are equations for the dependence of physical quantities on a physical scale, and do not depend on the factorization scale. The independence on the factorization scheme in the physical basis, as opposed to the conventional approach, would explicitly show up  when including terms that are one order higher in $\as$. 

The mathematical structure of Eqs.~(\ref{eq:F2dglap}) and (\ref{eq:FLdglap}) differ from the usual DGLAP equations for PDFs by the presence of derivatives of the structure functions on the r.h.s. To demonstrate that this slight difference   is not a problem in practice, we have also  solved them numerically using standard techniques. To be able to compare to existing DGLAP parametrizations, we have calculated  the initial conditions for the structure functions at $Q^2=2.0~\mathrm{GeV}^2$  through Eqs.~(\ref{eq:F2master}) and (\ref{eq:FLmaster}) at the first non-zero order in $\as$, using several available LO PDF sets including only light quarks i.e. PDFs in the 3-flavour scheme. The $Q^2$ evolution  using Eqs.~(\ref{eq:F2dglap}) and (\ref{eq:FLdglap})  is then performed by utilizing the ordinary differential equation (ODE) solver from the GNU Scientific Library~\cite{gough2009gnu}. The PDF sets used in this work are from the CTEQ (\texttt{CT14lo\_NF3}~\cite{Dulat:2015mca}), MSTW (\texttt{MSTW2008lo68cl\_nf3}~\cite{Martin:2010db}) and NNPDF (\texttt{NNPDF30\_lo\_as\_0118\_nf\_3}~\cite{NNPDF:2014otw}) collaborations, taken from the LHAPDF library~\cite{Buckley_2015}. In these leading order PDF sets, the scale-dependence of the quark-singlet and gluon PDFs obey the DGLAP equations in Eqs.~\eqref{eq:quarkdglap} and \eqref{eq:gdglap}
and the result of solving the $Q^2$ dependence through our Eqs.~(\ref{eq:F2dglap}) and (\ref{eq:FLdglap}) will, in fact, be identical to the one from using the evolved PDFs to compute $F_2$ and $F_{\rm L}$ via Eqs.~(\ref{eq:F2master}) and (\ref{eq:FLmaster}). Thus, at leading order, the comparison merely provides us a way to check our analytical results and their numerical implementation, while the scheme dependence of PDFs would only show up at higher order in $\as$. The computed $Q^2$ dependences of $\ft$ and $\fl$ are shown in Fig.~\ref{fig:results}. The results are shown separately for the three different PDF sets used to construct the initial condition for the $Q^2$ evolution. In the case of CT14 and MSTW2008 we find an excellent agreement between the results obtained by using the DGLAP evolved PDFs and by directly evolving the structure functions. For  the NNPDF3 PDF set the results clearly deviate in the  $x\lesssim 10^{-6}$ region. We believe, however, that this is a numerical problem with \texttt{NNPDF30\_lo\_as\_0118\_nf\_3}: we have checked that this particular set of PDFs also fails to satisfy the $Q^2$ independence of the momentum sum rule, 
\begin{align}
\frac{\dd}{\dd Q^2} \int_0^1 \dd{x} x \left[g(x,Q^2) + \Sigma(x,Q^2) \right] = 0 \,.
\end{align}
By substituting $g$ and $\Sigma$ from Eqs.~(\ref{eq:sigmafromF2}) and (\ref{eq:gfromF2FL}) we have checked that our implementation of the evolution does fulfill the momentum sum rule.

\section{Evolution of a six-observable physical basis}
\label{sec:DGLAPevolutioninphysicalbasiswithvalencequarks}

In this section we repeat the same steps as in Sec.~\ref{sec:DGLAPevolutioninphysicalbasis}, but now considering a more complete setup which corresponds to a full flavor separation of PDFs in three-flavour scheme. In this case the partonic degrees are the gluon, together with $u$, $\overline u$, $d$, $\overline d$ and $s=\overline s$ quark PDFs. We will thus need a set of six linearly independent DIS structure functions to set up a proper basis. As already in our simpler example, $\fl$ and $\ft$ give access to the gluons and quark singlet. This is complemented by the third neutral current structure function $\fk$, which is sensitive to the valence quarks. For flavor separation one needs charged current structure functions, where we choose $\ftw$, $\fkw$ to give access to positively and negatively charged (anti)quarks separately. Finally, to separate  strange from down quarks, we need a process that differentiates between quark generations in the final state. For this we choose  $\ftcw$, which corresponds to $\ftw$ with a tagged charm quark in the final state. At the leading CKM matrix element level this is only sensitive to $\overline{s}$.
Thus our basis of six structure functions consists of the neutral-current structure functions $\fl$, $\ft$ and $\fk$ plus the charged-current structure functions $\ftw$, $\fkw$, and $\ftcw$, corresponding to the $W^{-}$-boson exchange. 
To first non-zero order in $\as$, the expressions for these structure functions in terms of PDFs read~\cite{Furmanski:1981cw,ParticleDataGroup:2022pth,Moch:2004xu},
\begin{align}
    \label{eq:F2fullB}
    \ft \xq & = x \sum_q e_q^2 \Big[ q(x,Q^2)+\overline{q}(x,Q^2) \Big] \,, \\
    \label{eq:F3fullB}
    \fk \xq & = 2\sum_{q}(L^2_q-R_q^2) \Big[ q(x,Q^2) - \overline{q}(x,Q^2) \Big] \,, \\
    \label{eq:F2WfullB}
    \ftw \xq & = 2x \Big[ u(x,Q^2)+\overline{d}(x,Q^2)+\overline{s}(x,Q^2) \Big] \,, \\
    \label{eq:F3WfullB}
    \fkw \xq & = 2 \Big[ u(x,Q^2)-\overline{d}(x,Q^2)-\overline{s}(x,Q^2) \Big] \,, \\
    \label{eq:F2CWfullB}
    \ftcw \xq & = 2x\overline{s}\xq \,, \\
    \label{eq:FLfullB}
    \fl(x, Q^2) & = \frac{\as(Q^2)}{2\pi}2x\seq \int_{x}^{1}\frac{\dd{z}}{z}  \Bigg[ \cf z\left[q\left(\frac{x}{z}, Q^2\right)+\overline{q}\left(\frac{x}{z}, Q^2\right)\right] +4\TR z\left(1-z\right)g\left(\frac{x}{z}, Q^2\right) \Bigg]\\
    & \equiv \frac{\as(Q^2)}{2\pi}x\left[\cflwft^{(1)}\otimes\wft(Q^2) +2\nf\seqav\cflg^{(1)}\otimes g(Q^2)\right] \,, \nonumber
\end{align}
where we have already equated the renormalization and factorization scale with $Q^2$. Here, $L_q = T_q^3-2e_q\sin^2 \theta_W$ and $R_q =-2e_q\sin^2 \theta_W$, where $\theta_W$ denotes the Weinberg angle and $T_q^3$ is the third component of the weak isospin. 
The structure function coefficient functions $\cflwft^{(1)} =\cfls^{(1)}$, and $\cfls^{(1)}$ and $\cflg^{(1)}$ were defined in Eqs. \eqref{eq:C1FLS} and \eqref{eq:C1FLg}. 

Experimental constraints for most of these structure functions are available e.g. from the HERA collider~\cite{H1:2015ubc,H1:2013ktq,ZEUS:2014thn} and neutrino DIS experiments~\cite{Berge:1989hr,CHORUS:2005cpn, NuTeV:2005wsg,NuTeV:2001dfo,NOMAD:2013hbk}, and more are expected from future DIS experiments~\cite{AbdulKhalek:2021gbh,LHeC:2020van}. However, similarly to the case of global analyses of PDFs, further constraints e.g. from the LHC will presumably be ultimately required to have a good control over all of them.
Note that it is not necessary for the experimental data to be given precisely for one of the structure functions in our basis: because we have a full basis different  measured cross sections can be expressed in terms of the basis structure functions, such the reduced cross section in terms of $\fl$ and $\ft$.

We can write Eqs. \eqref{eq:F2fullB}--\eqref{eq:F2CWfullB} in a matrix form, 
\begin{align}
\left(\begin{array}{c} \ft \\
 \fk \\
 \ftw \\
 \fkw \\
 \ftcw \end{array} \right) & = 
 \left(
\begin{array}{ccccc}
x\ed & x\ed & x\eu & x\eu & 2x\es\\
2A_d & -2A_d & 2A_u & -2A_u &  0 \\
0 & 2x & 2x & 0 & 2x\\
0 & -2 & 2 & 0 & -2 \\
0 & 0 & 0 & 0 & 2x 
\end{array}
\right)
\left(
\begin{array}{c} d \\ \overline{d} \\ u \\ \overline{u} \\ \overline{s} \end{array} \right) ,
\end{align}
where we have defined $A_q \equiv L^2_q-R_q^2$ in order to simplify the notation. The determinant of the matrix on the right-hand side is non-zero, so it is invertible and we can express the quark PDFs in terms of $\ft$, $\fk$, $\ftw$, $\fkw$ and $\ftcw$: 
\begin{align}
    \label{eq:FullBd}
    xd\xq & = \frac{1}{A_u\ed+A_d\eu}\Bigg[ A_u\ft\xq+\frac{\eu}{2}x\fk\xq-\frac{A_u(2\eu+\ed)-A_d\eu}{4}\ftw\xq \\ 
    & -\frac{A_u(2\eu-\ed)+A_d\eu}{4}x\fkw\xq + \frac{A_u(\ed-2\es)-A_d\eu}{2}\ftcw\xq \Bigg] \,, \nonumber \\
    x\overline{d} \xq & = \frac{1}{4}\ftw\xq-\frac{1}{4}x\fkw\xq-\frac{1}{2}\ftcw \,, \\
    xu\xq & = \frac{1}{4}\ftw\xq+\frac{1}{4}x\fkw\xq \,, \\
    x\overline{u}\xq & = \frac{1}{A_u\ed+A_d\eu}\Bigg[ A_d\ft\xq-\frac{\ed}{2}x\fk\xq-\frac{A_d(2\ed+\eu)-A_u\ed}{4}\ftw\xq \\ 
    & + \frac{A_d(2\ed-\eu)+A_u\ed}{4}x\fkw\xq \Bigg] \,, \nonumber \\
    x\overline{s}\xq & = xs \xq =\frac{1}{2}\ftcw\xq \,. 
    \label{eq:FullBs}
\end{align}
For the gluon PDF we obtain the same expression as in two observable basis
\begin{align}
    \label{eq:g(x)fullB}
    g(x, Q^2) & = \int_x^1 \frac{\dd{z}}{z}\delta(1-z)\Bigg\{ \frac{\cf}{4 \TR\nf\seqav} \left[\frac{x}{z}\frac{\dd}{\dd\frac{x}{z}} -2 \right]\frac{\ft\left(\frac{x}{z}, Q^2\right)}{\frac{x}{z}}
            +\frac{1}{8\TR\nf\seqav}\left[ \frac{x^2}{z^2}\frac{\dd[2]}{\dd\left(\frac{x}{z}\right)^2}-2\frac{x}{z}\frac{\dd}{\dd\frac{x}{z}}+2 \right]\frac{\fl\left(\frac{x}{z}, Q^2\right)}{\frac{x}{z}\frac{\as(Q^2)}{2\pi}}\Bigg\} \nonumber \\
            & \equiv\frac{1}{\nf\seqav}\Bigg\{ \cgf_{g\wftp}\otimes \wftp +\cgf_{g\wft}\otimes \wft 
            +\cgf_{g\wflpp}\otimes\wflpp+\cgf_{g\wflp}\otimes\wflp+\cgf_{g\wfl}\otimes\wfl  \Bigg\},
\end{align}
where $\wfl, \wflpp$ and $\widetilde{F'}_{2,L}$ are given in Eqs.~\eqref{eq:wfl}--\eqref{eq:wflpp}, and the coefficient functions are  defined in Eqs. \eqref{eq:Cgwftp}--\eqref{eq:Cgwfl}. By taking derivatives of Eqs.~\eqref{eq:F2fullB}--\eqref{eq:FLfullB} with respect to $\log Q^2$ we arrive with evolution equations for the structure functions $\ft$, $\fk$, $\ftw$, $\fkw$, and $\ftcw$:
\begin{align}
    \label{eq:F2dglapFullB}
    \frac{\dd\ft(x, Q^2)}{\dd \log Q^2} & =  
    \frac{\as(Q^2)}{2\pi}x\left[P_{qq}\otimes \wft+2\seq P_{qg}\otimes g\right] \,, \\
    \label{eq:F3dglap}
    \frac{\dd \fk\xq}{\dd \log Q^2} & = 
    \frac{\as(Q^2)}{2\pi} P_{qq}\otimes\fk \,, \\
    \label{eq:F2Wdglap}
    \frac{\dd\ftw\xq}{\dd \log Q^2} & = 
    \frac{\as(Q^2)}{2\pi}x\left[P_{qq}\otimes \wftw+6 P_{qg}\otimes g\right] \,, \\
    \label{eq:F3Wdglap}
    \frac{\dd \fkw\xq}{\dd \log Q^2} & = 
    \frac{\as(Q^2)}{2\pi}\left[P_{qq}\otimes \fkw -2 P_{qg}\otimes g\right] \,, \\
    \label{eq:F2CWdglap}
    \frac{\dd\ftcw\xq}{\dd \log Q^2} & =
    \frac{\as(Q^2)}{2\pi}x\left[P_{qq}\otimes \wftcw+2 P_{qg}\otimes g\right] \,,
    \end{align}
where $\wftw\xq \equiv \ftw\xq/x$ and $\wftcw\xq \equiv \ftcw\xq/x$. 
Here the convolution $P_{qg}\otimes g$ appearing in Eqs.~\eqref{eq:F2dglapFullB}--\eqref{eq:F2CWdglap} can be given in a single integral form by using Eq.~\eqref{eq:g(x)fullB} 
\begin{align}
    \label{eq:intPqg}
   P_{qg}\otimes g =   \frac{1}{4\nf\seqav}\Bigg\{&\cf\left[-\wft\xq+2\int_x^1\frac{\dd\xi}{\xi}\left(\frac{x}{\xi}-1\right)\wft(\xi, Q^2)\right] \nonumber \\
    +&\frac{1}{2}\left[ \wfl\xq-\wflp\xq+2\int_x^1\frac{\dd{\xi}}{\xi}\wfl(\xi, Q^2)\right]\Bigg\}\,.
\end{align}

In a similar way, we can obtain evolution equation for the structure function $\fl$: 
  \begin{align}
 \label{eq:FLdglapkonvoFullB}
    \frac{\dd}{\dd \log Q^2}\left(\frac{ \fl(x, Q^2)}{\frac{\as(Q^2)}{2\pi}}\right) & =  
    \frac{\as(Q^2)}{2\pi}x\Bigg\{2\left[\cflwft^{(1)}\otimes P_{qg}+\cflg^{(1)}\otimes P_{gg} \right]\otimes \left[\cgf_{g\wflpp}\otimes\wflpp+\cgf_{g\wflp}\otimes\wflp+\cgf_{g\wfl }\otimes\wfl \right] \nonumber \\
    & +\left[\cflwft^{(1)}\otimes\left(P_{qq}+2P_{qg}\otimes \cgf_{g\wft}\right)+2\cflg^{(1)}\otimes P_{gg}\otimes\cgf_{g\wft} \right]\otimes\wft \\
    & +2\left[\cflwft^{(1)}\otimes P_{qg}+\cflg^{(1)}\otimes P_{gg} \right]\otimes\cgf_{g\wftp}\otimes \wftp \nonumber \\
    & +
     \frac{\nf\seqav}{A_u\ed+A_d\eu}\cflg^{(1)}\otimes P_{gq}\otimes\Big[2\left(A_d+A_u \right)\wft +\left(\eu-\ed\right)\fk +\left( A_u-A_d\right)\left(\ed-\eu\right)\wftw  \nonumber \\
    & +\left( A_u+A_d\right)\left(\ed-\eu\right)\fkw  \Big]  \Bigg\}. \nonumber 
\end{align}
  Opening the convolutions in Eq.~\eqref{eq:FLdglapkonvoFullB} and performing a partial integration we can write the DGLAP evolution of $\fl\xq / \frac{\as}{2\pi}$ in a single integral form as

\begin{align}
    \label{eq:FLdglapFullB}
    \frac{\dd}{\dd \log Q^2}\left(\frac{ \fl(x, Q^2)}{\frac{\as(Q^2)}{2\pi}}\right) =  \frac{\as(Q^2)}{2\pi}x&\Big\{ P_{\fl \wft}\otimes\wft +P_{\fl \wfl}\otimes\wfl +P_{\fl \wflp}\otimes\wflp + P_{\fl \fk}\otimes\fk  \nonumber \\
    & + P_{\fl \wftw}\otimes\wftw+ P_{\fl \fkw}\otimes\fkw \Big\} \,.
\end{align}
By defining a function
\begin{equation}
    \label{eq:fullPFLsigma}
    \rho(z) \equiv \frac{\nf\seqav}{A_u\ed+A_d\eu}\cflg^{(1)}\otimes P_{gq} = \frac{4\TR\cf \nf\seqav}{A_u\ed+A_d\eu}\left[  -1+\frac{1}{3z}+\frac{2}{3}z^2-z\log\left(z\right)\right],
\end{equation}
we can write the kernels as
\begin{align}
    \label{eq:fullPFLF2}
    P_{\fl \wft}(z) & \equiv \cf \Bigg[4\nc-2\nc\frac{1}{z}+2\left(\cf+\nc-b_0\right)z-4\nc z^2 \nonumber \\ 
    &+4\left(2\nc -\cf\right)z\log\left(z\right)+4\left(\cf-\nc\right)z\log\left(1-z\right) \Bigg] + 2\left(A_d+A_u \right)\rho(z)\,,\\
    \label{eq:fullPFLF3}
    P_{\fl \fk}(z) & \equiv  \left(\eu-\ed\right)\rho(z)\,,\\
    \label{eq:fullPFLF2w}
    P_{\fl \wftw}(z) & \equiv \left( A_u-A_d\right)\left(\ed-\eu\right)\rho(z)\,,\\
    \label{eq:fullPFLF3w}
    P_{\fl \fkw}(z) & \equiv \left( A_u+A_d\right)\left(\ed-\eu\right)\rho(z)\,.
\end{align}
Here the kernels $P_{\fl \wfl} $ and $P_{\fl \wflp}$ are the same as in the case of two-observable physical basis, $P_{\fl \wfl}^{2\text{-}o.}$ and $P_{\fl \wflp}^{2\text{-}o.}$ respectively, in Eqs.~\eqref{eq:PFLFL} and \eqref{eq:PFLdFL}.

\begin{figure}[htb]
  \includegraphics[width=.5\linewidth]{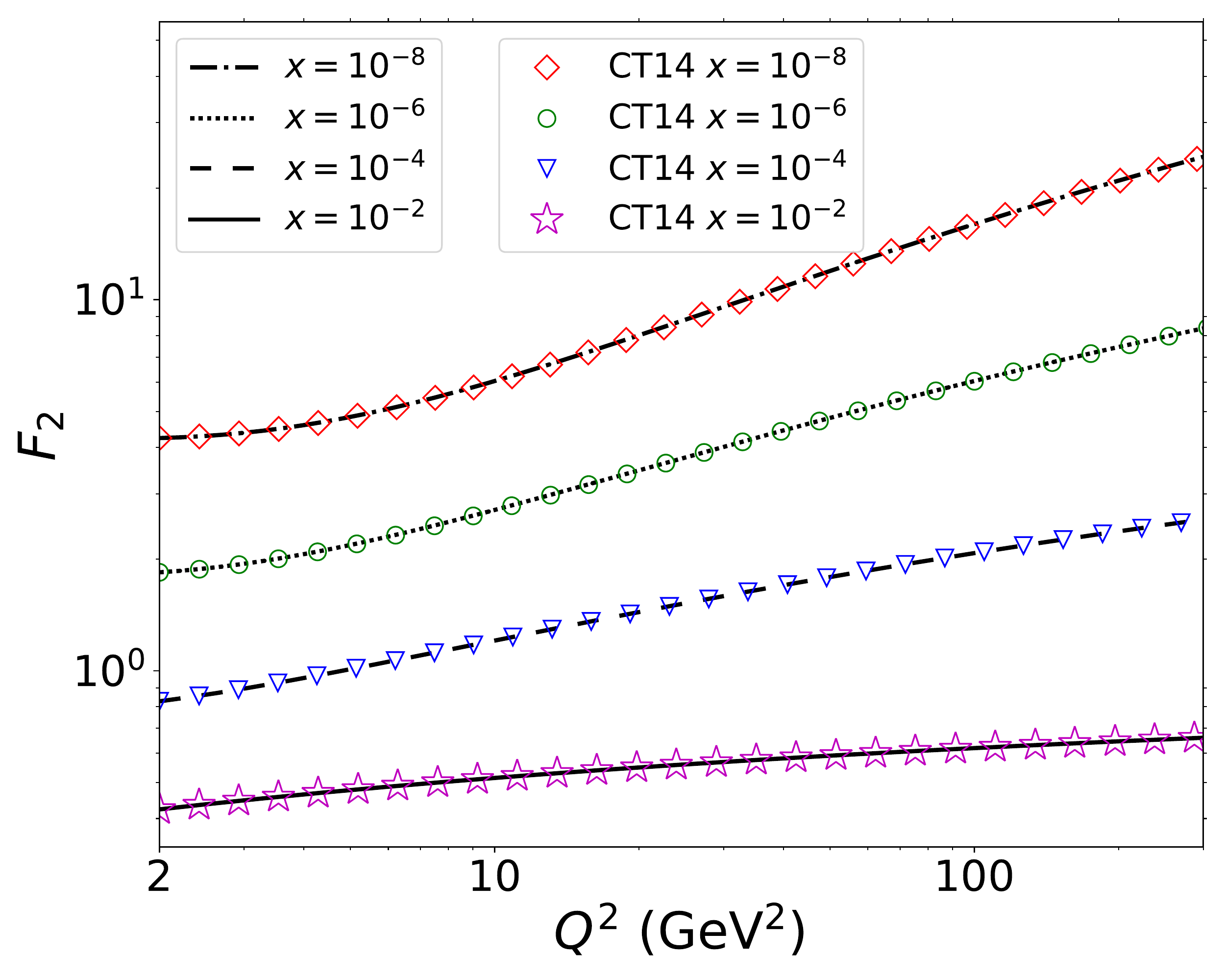}\hfill
  \includegraphics[width=.5\linewidth]{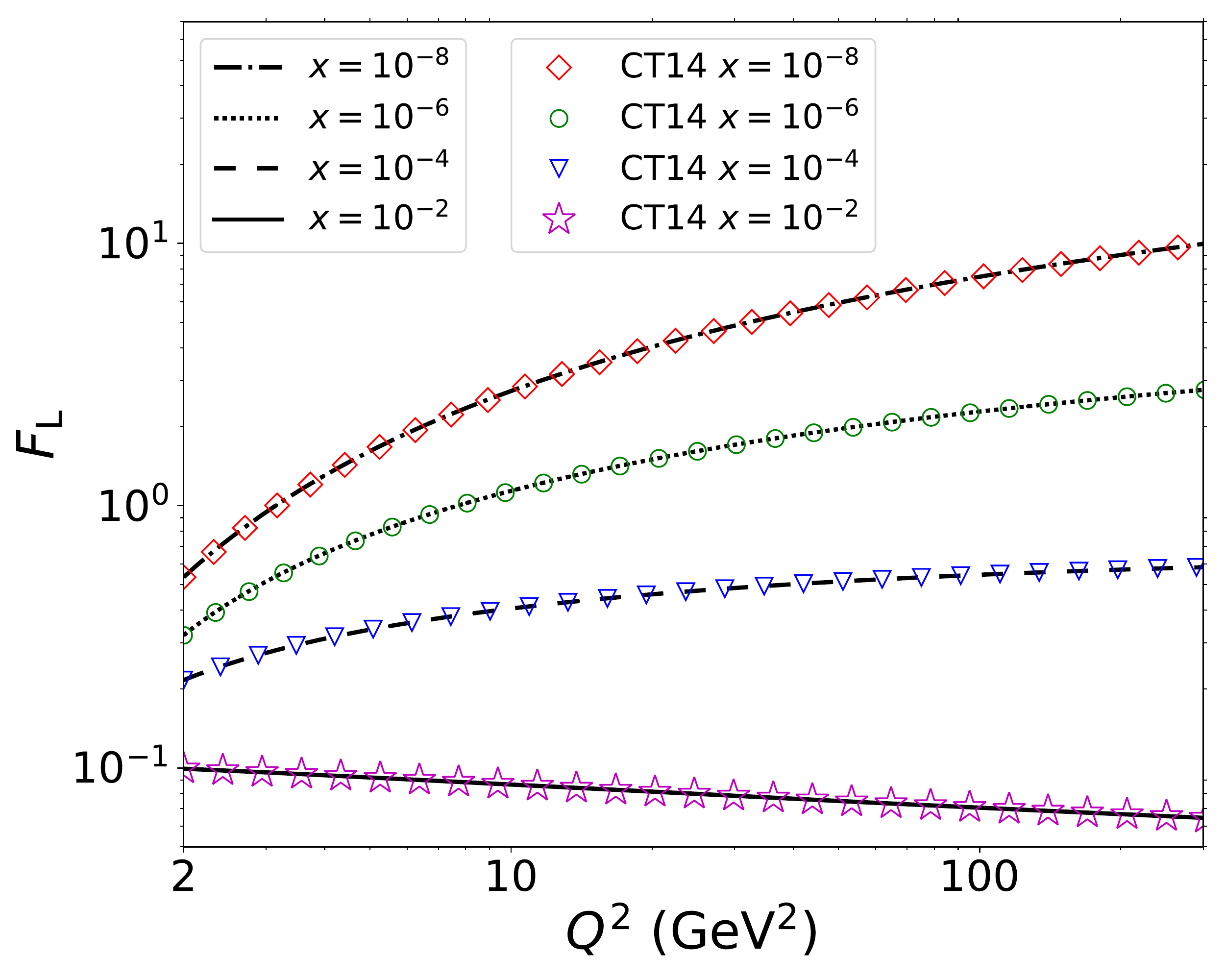}
  \includegraphics[width=.5\linewidth]{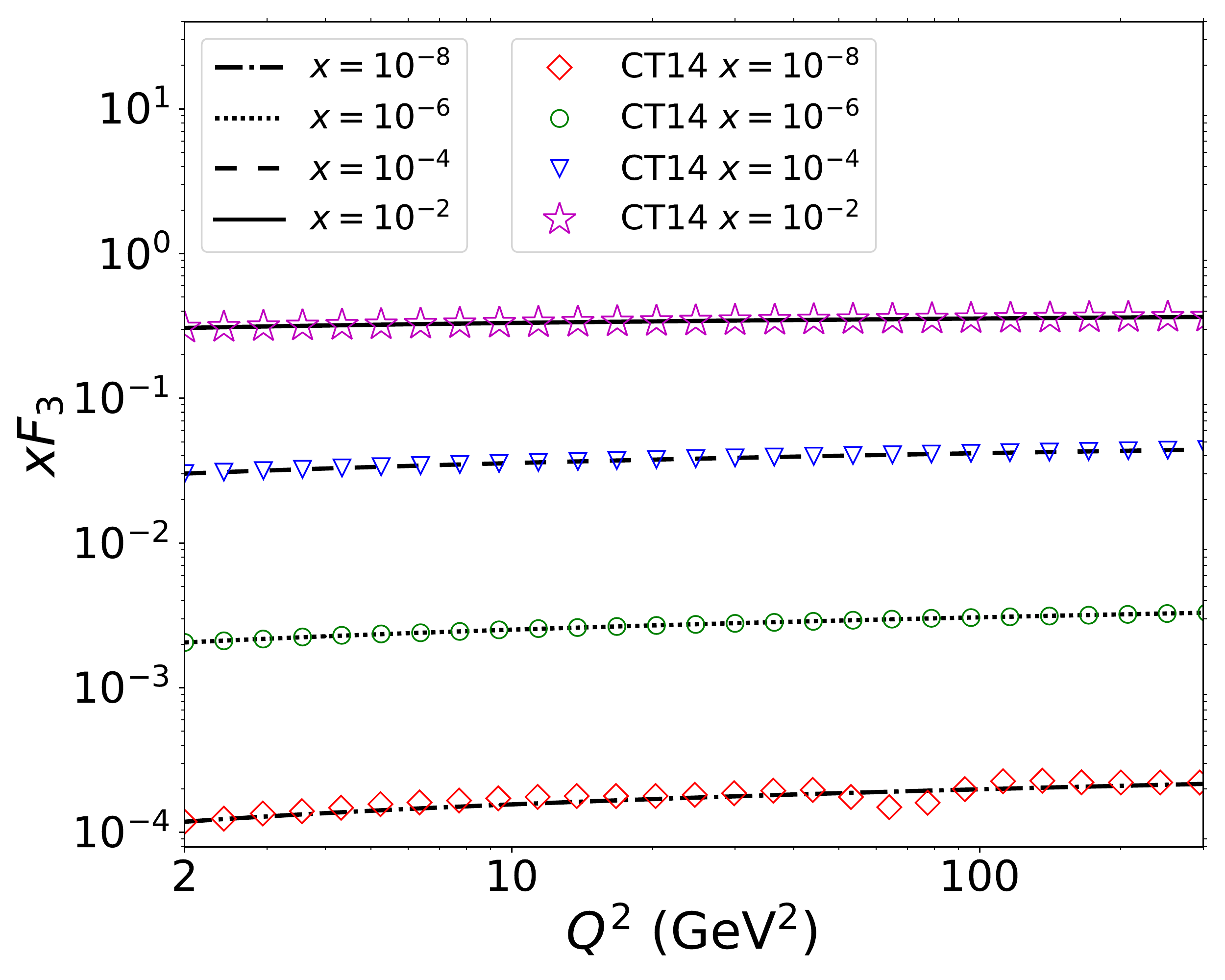}\hfill
  \includegraphics[width=.5\linewidth]{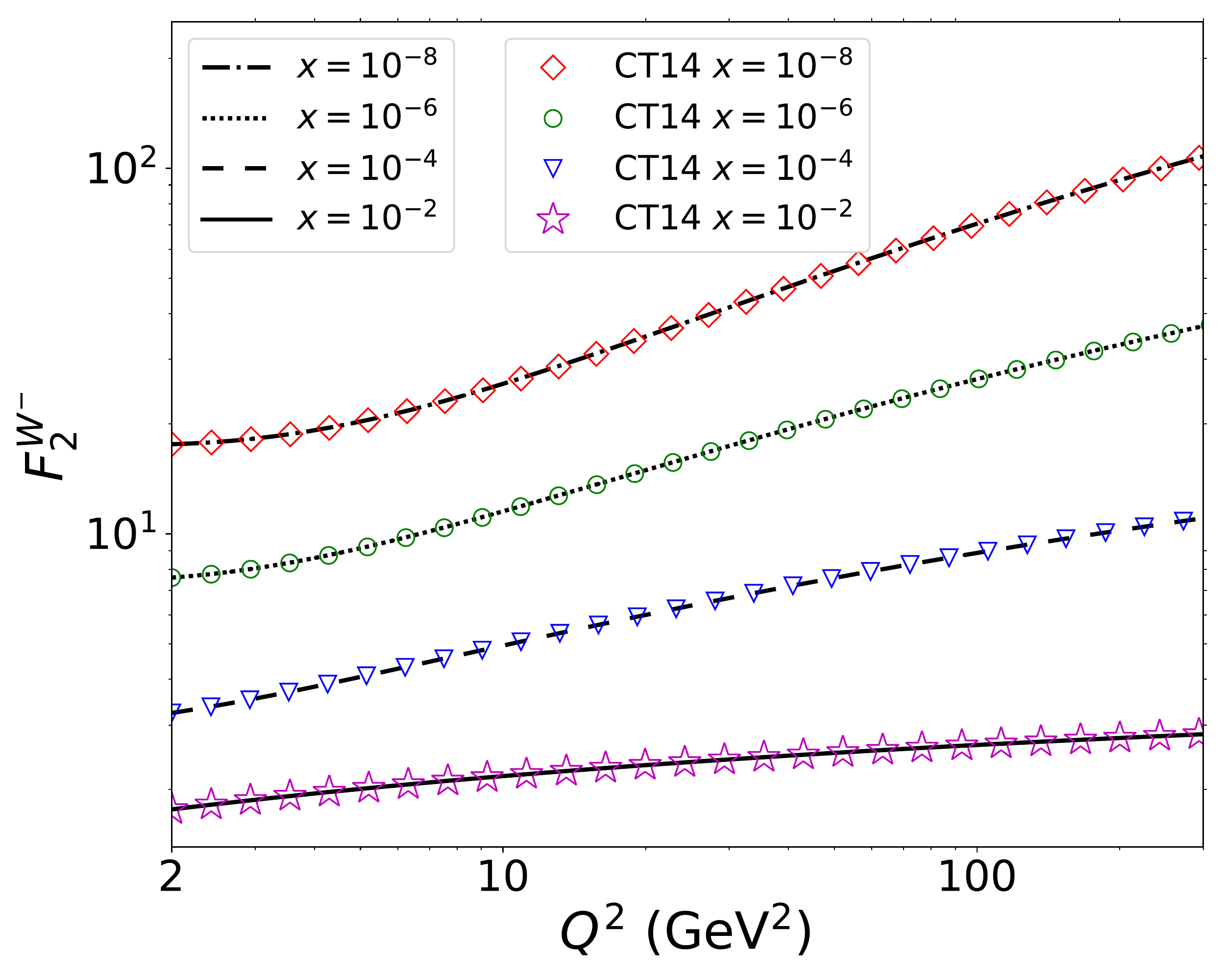}
  \includegraphics[width=.5\linewidth]{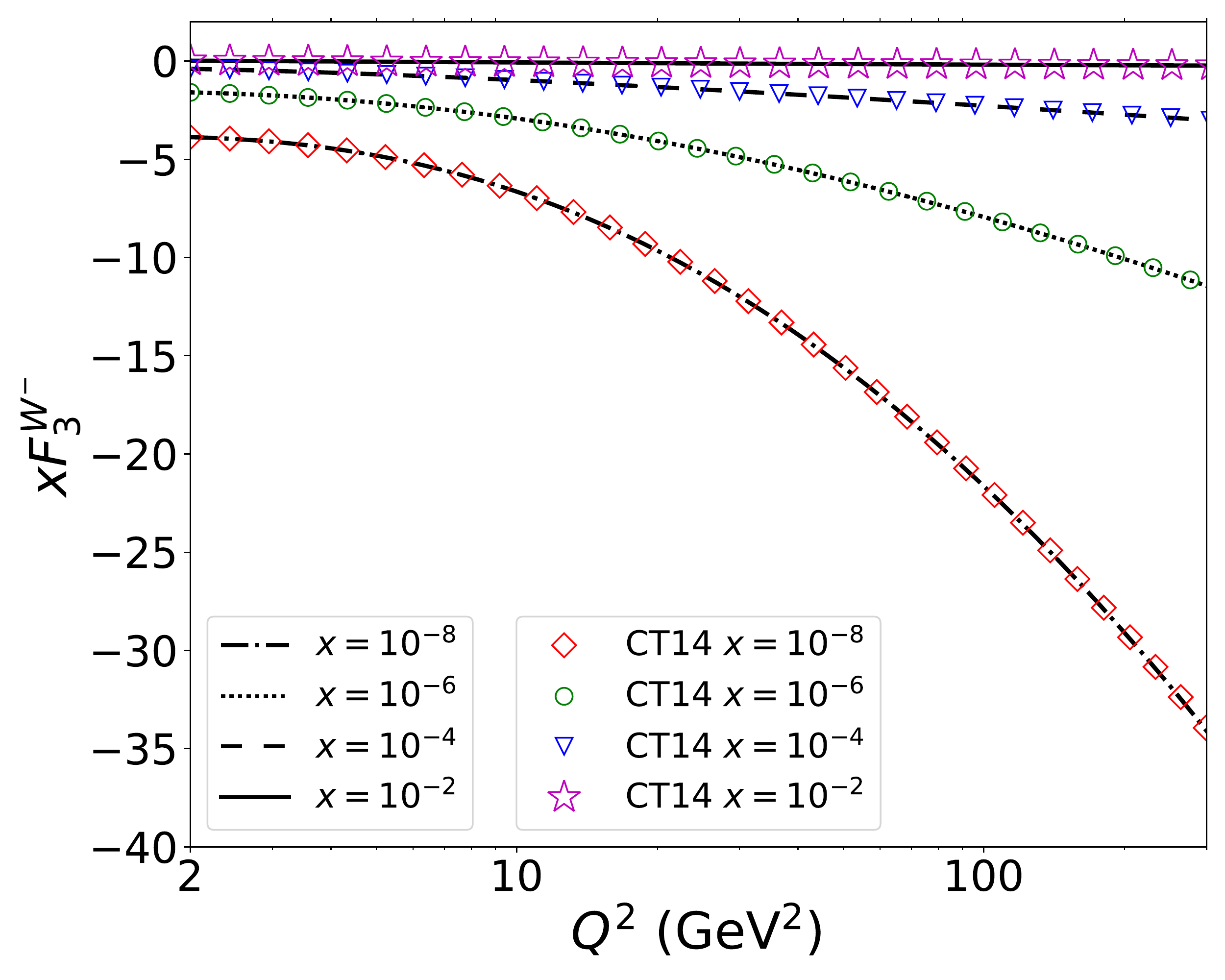}\hfill
  \includegraphics[width=.5\linewidth]{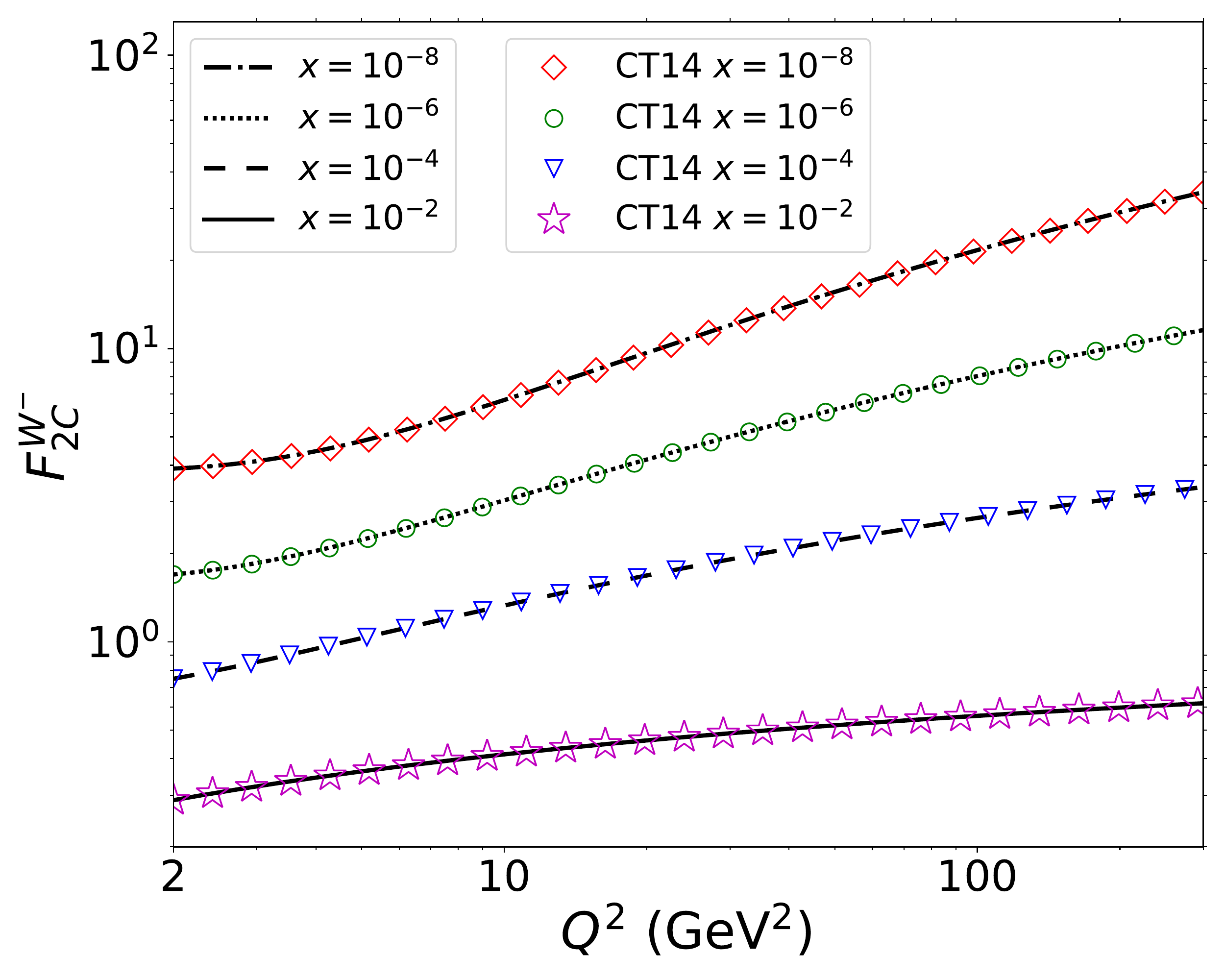}
  \caption{
  The $Q^2$ dependence of $\fl$, $\ft$, $\fk$, $\ftw$, $\fkw$, and $\ftcw$ using the physical-basis approach (curves) compared with the usual PDF-based approach (markers).}
\label{fig:resultsFullBasis}
\end{figure}

We have solved the set of evolution equations in Eqs.~\eqref{eq:F2dglapFullB}--\eqref{eq:FLdglapkonvoFullB} using the same methods as in Sec.~\ref{sec:DGLAPevolutioninphysicalbasis}. This time we have set the initial condition by calculating the structure functions in Eqs.~\eqref{eq:F2fullB}--\eqref{eq:FLfullB} at $Q^2 = 2.0 \, {\rm GeV}^2$  using the \texttt{CT14lo\_NF3} set of LO PDFs. Also in the present case we can verify that the results obtained by solving Eqs.~\eqref{eq:F2dglapFullB}--\eqref{eq:FLdglapkonvoFullB} match those obtained by using the usual PDF-based approach  where the structure functions in Eqs.~\eqref{eq:F2fullB}--\eqref{eq:FLfullB} are evaluated using DGLAP-evolved PDFs at higher scale. The results are shown in Fig.~\ref{fig:resultsFullBasis}, and the two approaches are again found to agree within the numerical accuracy. The only visible differences are in $\fk$, defined in Eq. \eqref{eq:F3fullB}, at small $x$. As $\fk$ is proportional to valence quarks, it tends to zero at small $x$ and the wiggles seen in Fig.~\ref{fig:resultsFullBasis} in the results computed directly with \texttt{CT14lo\_NF3} are presumably a result of numerical instability associated with small numbers.

\section{Conclusions and Outlook}
\label{sec:conclusions}

We have in this paper advocated an approach to the $Q^2$ dependence of DIS structure functions where the DGLAP evolution equations are formulated and solved directly for measurable structure functions. The evolution equations in the physical basis are more complicated than when the $Q^2$ dependence is formulated in terms of PDFs, and also e.g. derivatives of the structure functions are involved. However, this feature is compensated by several advantages. Most importantly, there is no need to define the factorization scheme nor the factorization scale: since the evolution equations express the $Q^2$-dependence of observable quantities in terms of observable quantities, the coefficients must necessarily be scheme and scale independent. 
The approach in terms of physical structure functions also has the advantage of being more transparent in the parametrization of the initial conditions of the evolution, and does not have ambiguities concerning  positivity  constraints.

We have shown that it is possible to find a basis of physical structure functions that has a one-to-one correspondence with PDFs in the 3-flavour scheme, which leads to a closed set of coupled evolution equations among the physical structure functions. We formulate the evolution equations directly in momentum space, in contrast to the existing literature on the subject where the evolution has been written in  Mellin space. At this stage we have presented the result only to the first non-zero order in $\as$ but the approach can be straightforwardly extended to the second non-zero order as well, which we intend to do in a future work. 
The qualitative novelty at second non-zero order is that also double derivatives of structure functions will be involved which can potentially cause extra challenges for a precise numerical evaluation. The emergence of higher derivatives is easy to understand through a simple example: Suppose we have a structure function $F$ that depends on the gluon PDF through,
\begin{equation}
F \equiv C^{(1)} \otimes g + \alpha_s C^{(2)} \otimes g  \,, 
\end{equation}
and a differential operator $P$ such that $P C^{(1)} \otimes g = g$. Then,
\begin{align}
g = P \left[F - \alpha_s C^{(2)} \otimes g \right] = P \left[F - \alpha_s C^{(2)} \otimes PF \right] + \mathcal{O}(\alpha_s^2) \,.
\end{align}
If $P$ contains a derivative, the second term will contain a double derivative. At still higher orders in $\alpha_s$ even higher derivatives will appear. Another complication that can be expected at higher orders is that the derivation of the evolution kernels may involve integrals which cannot be analytically performed and have to be dealt with in a different way. In the lowest order discussed here it was still possible to express all the integrals in a closed form e.g. in going from Eq.~(\ref{eq:toyfleq}) to Eq.~(\ref{eq:FLdglap}). The actual scheme dependence in the conventional PDF approach will appear only at the second non-zero order. Thus in our leading order calculation we have been able to test numerically our evolution equations against results from standard PDF sets. 

For simplicity, we have here worked in terms of 3-flavour basis whose applicability is restricted to ``low'' $Q^2$ due to appearance of logarithmically enhanced contributions involving the heavy-quark masses, $\alpha_s^m\log^n(m^2/Q^2), \ m \geq n$, at NLO and beyond. In terms of PDFs, these logarithms are resummed into the scale-dependent heavy-quark PDFs. To accomplish the same in the physical-basis approach, one needs to introduce additional heavy-quark dependent structure functions, like the neutral-current $F_{2c}$ and $F_{2b}$. Indeed, $F_{2c}$ is a significant fraction of e.g. the HERA total cross section. Each new flavour will increase the dimensionality of the evolution equations by one, but in a picture in which the heavy-quark PDFs are perturbatively generated, the initial conditions of the evolution are still determined by six structure functions. Including nonperturbative heavy flavor then requires additional initial conditions.
Additionally, if one includes explicit quark masses, the coefficient functions contain $Q^2 / m_q^2$ terms  which will add technical difficulties.
Also the CKM-suppressed contributions, neglected here for simplicity, can straightforwardly be included.

Since the PDFs can be expressed in terms of physical structure functions, it follows that all other cross sections -- those at the LHC in particular -- can be expressed in terms of structure functions as well. In other words, the measured structure functions could, in principle, be directly used to predict cross sections at the LHC without any reference to PDFs, as discussed in Sec.~\ref{sec:intro}. This may eventually provide an alternative approach to the usual PDF-based paradigm and a way to estimate theoretical uncertainties associated with the scheme-dependence of factorization.

\begin{acknowledgments}
This work was supported by the Academy of Finland, the Centre of Excellence in Quark Matter (projects 346324 and 346326), projects 338263 and 346567 (H.M), projects 321840 (T.L, M.T), and project 308301 (H.P., M.T). This work was also supported under the European Union’s Horizon 2020 research and innovation programme by the European Research Council (ERC, grant agreement No. ERC-2018-ADG-835105 YoctoLHC) and by the STRONG-2020 project (grant agreement No. 824093). The content of this article does not reflect the official opinion of the European Union and responsibility for the information and views expressed therein lies entirely with the authors. 
\end{acknowledgments}

\bibliographystyle{JHEP-2modlong.bst}
\bibliography{refs}

\providecommand{\href}[2]{#2}\begingroup\raggedright\begin{thebibliography}{10}

\bibitem{Collins:1989gx}
J.~C. Collins, D.~E. Soper and G.~F. Sterman, {\it {Factorization of Hard
  Processes in QCD}},  \href{http://dx.doi.org/10.1142/9789814503266_0001}{{\em
  Adv. Ser. Direct. High Energy Phys.} {\bf 5} (1989) 1}
  [\href{http://arXiv.org/abs/hep-ph/0409313}{{\tt arXiv:hep-ph/0409313}}].

\bibitem{Gribov:1972ri}
V.~N. Gribov and L.~N. Lipatov, {\it {Deep inelastic e p scattering in
  perturbation theory}},  {\em Sov. J. Nucl. Phys.} {\bf 15} (1972) 438.

\bibitem{Lipatov:1974qm}
L.~N. Lipatov, {\it {The parton model and perturbation theory}},  {\em Yad.
  Fiz.} {\bf 20} (1974) 181.

\bibitem{Altarelli:1977zs}
G.~Altarelli and G.~Parisi, {\it {Asymptotic Freedom in Parton Language}},
  \href{http://dx.doi.org/10.1016/0550-3213(77)90384-4}{{\em Nucl. Phys. B}
  {\bf 126} (1977) 298}.

\bibitem{Dokshitzer:1977sg}
Y.~L. Dokshitzer, {\it {Calculation of the Structure Functions for Deep
  Inelastic Scattering and $e^+ e^-$ Annihilation by Perturbation Theory in
  Quantum Chromodynamics.}},  {\em Sov. Phys. JETP} {\bf 46} (1977) 641.

\bibitem{Ethier:2020way}
J.~J. Ethier and E.~R. Nocera, {\it {Parton Distributions in Nucleons and
  Nuclei}},  \href{http://dx.doi.org/10.1146/annurev-nucl-011720-042725}{{\em
  Ann. Rev. Nucl. Part. Sci.} {\bf 70} (2020) 43}
  [\href{http://arXiv.org/abs/2001.07722}{{\tt arXiv:2001.07722 [hep-ph]}}].

\bibitem{Bardeen:1978yd}
W.~A. Bardeen, A.~J. Buras, D.~W. Duke and T.~Muta, {\it {Deep Inelastic
  Scattering Beyond the Leading Order in Asymptotically Free Gauge Theories}},
  \href{http://dx.doi.org/10.1103/PhysRevD.18.3998}{{\em Phys. Rev. D} {\bf 18}
  (1978) 3998}.

\bibitem{Altarelli:1998gn}
G.~Altarelli, S.~Forte and G.~Ridolfi, {\it {On positivity of parton
  distributions}},  \href{http://dx.doi.org/10.1016/S0550-3213(98)00661-0}{{\em
  Nucl. Phys. B} {\bf 534} (1998) 277}
  [\href{http://arXiv.org/abs/hep-ph/9806345}{{\tt arXiv:hep-ph/9806345}}].

\bibitem{Candido:2020yat}
A.~Candido, S.~Forte and F.~Hekhorn, {\it {Can $ \overline{\mathrm{MS}} $
  parton distributions be negative?}},
  \href{http://dx.doi.org/10.1007/JHEP11(2020)129}{{\em JHEP} {\bf 11} (2020)
  129} [\href{http://arXiv.org/abs/2006.07377}{{\tt arXiv:2006.07377
  [hep-ph]}}].

\bibitem{Collins:2021vke}
J.~Collins, T.~C. Rogers and N.~Sato, {\it {Positivity and renormalization of
  parton densities}},
  \href{http://dx.doi.org/10.1103/PhysRevD.105.076010}{{\em Phys. Rev. D} {\bf
  105} (2022)~no.~7 076010} [\href{http://arXiv.org/abs/2111.01170}{{\tt
  arXiv:2111.01170 [hep-ph]}}].

\bibitem{Candido:2023ujx}
A.~Candido, S.~Forte, T.~Giani and F.~Hekhorn, {\it {On the positivity of MSbar
  parton distributions}},  \href{http://arXiv.org/abs/2308.00025}{{\tt
  arXiv:2308.00025 [hep-ph]}}.

\bibitem{Armesto:2022mxy}
N.~Armesto, T.~Lappi, H.~M\"antysaari, H.~Paukkunen and M.~Tevio, {\it
  {Signatures of gluon saturation from structure-function measurements}},
  \href{http://dx.doi.org/10.1103/PhysRevD.105.114017}{{\em Phys. Rev. D} {\bf
  105} (2022)~no.~11 114017} [\href{http://arXiv.org/abs/2203.05846}{{\tt
  arXiv:2203.05846 [hep-ph]}}].

\bibitem{vanNeerven:1999ca}
W.~L. van Neerven and A.~Vogt, {\it {NNLO evolution of deep inelastic structure
  functions: The Nonsinglet case}},
  \href{http://dx.doi.org/10.1016/S0550-3213(99)00668-9}{{\em Nucl. Phys. B}
  {\bf 568} (2000) 263} [\href{http://arXiv.org/abs/hep-ph/9907472}{{\tt
  arXiv:hep-ph/9907472}}].

\bibitem{Ball:2017otu}
R.~D. Ball, V.~Bertone, M.~Bonvini, S.~Marzani, J.~Rojo and L.~Rottoli, {\it
  {Parton distributions with small-x resummation: evidence for BFKL dynamics in
  HERA data}},  \href{http://dx.doi.org/10.1140/epjc/s10052-018-5774-4}{{\em
  Eur. Phys. J. C} {\bf 78} (2018)~no.~4 321}
  [\href{http://arXiv.org/abs/1710.05935}{{\tt arXiv:1710.05935 [hep-ph]}}].

\bibitem{xFitterDevelopersTeam:2018hym}
{\bf xFitter Developers' Team} collaboration, H.~Abdolmaleki {\em et.~al.},
  {\it {Impact of low-$x$ resummation on QCD analysis of HERA data}},
  \href{http://dx.doi.org/10.1140/epjc/s10052-018-6090-8}{{\em Eur. Phys. J. C}
  {\bf 78} (2018)~no.~8 621} [\href{http://arXiv.org/abs/1802.00064}{{\tt
  arXiv:1802.00064 [hep-ph]}}].

\bibitem{Hou:2019efy}
T.-J. Hou {\em et.~al.}, {\it {New CTEQ global analysis of quantum
  chromodynamics with high-precision data from the LHC}},
  \href{http://dx.doi.org/10.1103/PhysRevD.103.014013}{{\em Phys. Rev. D} {\bf
  103} (2021)~no.~1 014013} [\href{http://arXiv.org/abs/1912.10053}{{\tt
  arXiv:1912.10053 [hep-ph]}}].

\bibitem{LHCHiggsCrossSectionWorkingGroup:2013rie}
{\bf LHC Higgs Cross Section Working Group} collaboration, J.~R. Andersen {\em
  et.~al.}, {\it {Handbook of LHC Higgs Cross Sections: 3. Higgs Properties}},
  \href{http://arXiv.org/abs/1307.1347}{{\tt arXiv:1307.1347 [hep-ph]}}.

\bibitem{Harland-Lang:2018bxd}
L.~A. Harland-Lang and R.~S. Thorne, {\it {On the Consistent Use of Scale
  Variations in PDF Fits and Predictions}},
  \href{http://dx.doi.org/10.1140/epjc/s10052-019-6731-6}{{\em Eur. Phys. J. C}
  {\bf 79} (2019)~no.~3 225} [\href{http://arXiv.org/abs/1811.08434}{{\tt
  arXiv:1811.08434 [hep-ph]}}].

\bibitem{Furmanski:1981cw}
W.~Furmanski and R.~Petronzio, {\it {Lepton - Hadron Processes Beyond Leading
  Order in Quantum Chromodynamics}},
  \href{http://dx.doi.org/10.1007/BF01578280}{{\em Z. Phys. C} {\bf 11} (1982)
  293}.

\bibitem{Catani:1996sc}
S.~Catani, {\it {Physical anomalous dimensions at small x}},
  \href{http://dx.doi.org/10.1007/s002880050512}{{\em Z. Phys. C} {\bf 75}
  (1997) 665} [\href{http://arXiv.org/abs/hep-ph/9609263}{{\tt
  arXiv:hep-ph/9609263}}].

\bibitem{Blumlein:2000wh}
J.~Blumlein, V.~Ravindran and W.~L. van Neerven, {\it {On the Drell-Levy-Yan
  relation to {$\mathcal{O}(\as^2)$}}},
  \href{http://dx.doi.org/10.1016/S0550-3213(00)00422-3}{{\em Nucl. Phys. B}
  {\bf 586} (2000) 349} [\href{http://arXiv.org/abs/hep-ph/0004172}{{\tt
  arXiv:hep-ph/0004172}}].

\bibitem{Hentschinski:2013zaa}
M.~Hentschinski and M.~Stratmann, {\it {On the Practical Application of
  Physical Anomalous Dimensions}},  \href{http://arXiv.org/abs/1311.2825}{{\tt
  arXiv:1311.2825 [hep-ph]}}.

\bibitem{Blumlein:2021lmf}
J.~Bl\"umlein and M.~Saragnese, {\it {The N$^3$LO scheme-invariant QCD
  evolution of the non-singlet structure functions $F_2^\mathrm{NS}(x,Q^2)$ and
  $g_1^\mathrm{NS}(x,Q^2)$}},
  \href{http://dx.doi.org/10.1016/j.physletb.2021.136589}{{\em Phys. Lett. B}
  {\bf 820} (2021) 136589} [\href{http://arXiv.org/abs/2107.01293}{{\tt
  arXiv:2107.01293 [hep-ph]}}].

\bibitem{RuizArriola:1998er}
E.~Ruiz~Arriola, {\it {NLO evolution for large scale distances, positivity
  constraints and the low-energy model of the nucleon}},
  \href{http://dx.doi.org/10.1016/S0375-9474(98)00489-8}{{\em Nucl. Phys. A}
  {\bf 641} (1998) 461}.

\bibitem{Baulieu:1979mr}
L.~Baulieu, E.~G. Floratos and C.~Kounnas, {\it {Parton Model Interpretation of
  the Cut Vertex Formalism}},
  \href{http://dx.doi.org/10.1016/0550-3213(80)90230-8}{{\em Nucl. Phys. B}
  {\bf 166} (1980) 321}.

\bibitem{Floratos:1980hm}
E.~G. Floratos, R.~Lacaze and C.~Kounnas, {\it {Space and Timelike Cut Vertices
  in {QCD} Beyond the Leading Order. 2. The Singlet Sector}},
  \href{http://dx.doi.org/10.1016/0370-2693(81)90016-2}{{\em Phys. Lett. B}
  {\bf 98} (1981) 285}.

\bibitem{Blumlein:2012bf}
J.~Blumlein, {\it {The Theory of Deeply Inelastic Scattering}},
  \href{http://dx.doi.org/10.1016/j.ppnp.2012.09.006}{{\em Prog. Part. Nucl.
  Phys.} {\bf 69} (2013) 28} [\href{http://arXiv.org/abs/1208.6087}{{\tt
  arXiv:1208.6087 [hep-ph]}}].

\bibitem{gough2009gnu}
B.~Gough, {\em GNU scientific library reference manual}.
\newblock Network Theory Ltd., 2009.

\bibitem{Dulat:2015mca}
S.~Dulat, T.-J. Hou, J.~Gao, M.~Guzzi, J.~Huston, P.~Nadolsky, J.~Pumplin,
  C.~Schmidt, D.~Stump and C.~P. Yuan, {\it {New parton distribution functions
  from a global analysis of quantum chromodynamics}},
  \href{http://dx.doi.org/10.1103/PhysRevD.93.033006}{{\em Phys. Rev. D} {\bf
  93} (2016)~no.~3 033006} [\href{http://arXiv.org/abs/1506.07443}{{\tt
  arXiv:1506.07443 [hep-ph]}}].

\bibitem{Martin:2010db}
A.~D. Martin, W.~J. Stirling, R.~S. Thorne and G.~Watt, {\it {Heavy-quark mass
  dependence in global PDF analyses and 3- and 4-flavour parton
  distributions}},
  \href{http://dx.doi.org/10.1140/epjc/s10052-010-1462-8}{{\em Eur. Phys. J. C}
  {\bf 70} (2010) 51} [\href{http://arXiv.org/abs/1007.2624}{{\tt
  arXiv:1007.2624 [hep-ph]}}].

\bibitem{NNPDF:2014otw}
{\bf NNPDF} collaboration, R.~D. Ball {\em et.~al.}, {\it {Parton distributions
  for the LHC Run II}},  \href{http://dx.doi.org/10.1007/JHEP04(2015)040}{{\em
  JHEP} {\bf 04} (2015) 040} [\href{http://arXiv.org/abs/1410.8849}{{\tt
  arXiv:1410.8849 [hep-ph]}}].

\bibitem{Buckley_2015}
A.~Buckley, J.~Ferrando, S.~Lloyd, K.~Nordström, B.~Page, M.~Rüfenacht,
  M.~Schönherr and G.~Watt, {\it {LHAPDF}6: parton density access in the {LHC}
  precision era},  \href{http://dx.doi.org/10.1140/epjc/s10052-015-3318-8}{{\em
  The European Physical Journal C} {\bf 75} (mar, 2015) }.

\bibitem{ParticleDataGroup:2022pth}
{\bf Particle Data Group} collaboration, R.~L. Workman {\em et.~al.}, {\it
  {Review of Particle Physics}},
  \href{http://dx.doi.org/10.1093/ptep/ptac097}{{\em PTEP} {\bf 2022} (2022)
  083C01}.

\bibitem{Moch:2004xu}
S.~Moch, J.~A.~M. Vermaseren and A.~Vogt, {\it {The Longitudinal structure
  function at the third order}},
  \href{http://dx.doi.org/10.1016/j.physletb.2004.11.063}{{\em Phys. Lett. B}
  {\bf 606} (2005) 123} [\href{http://arXiv.org/abs/hep-ph/0411112}{{\tt
  arXiv:hep-ph/0411112}}].

\bibitem{H1:2015ubc}
{\bf H1, ZEUS} collaboration, H.~Abramowicz {\em et.~al.}, {\it {Combination of
  measurements of inclusive deep inelastic ${e^{\pm }p}$ scattering cross
  sections and QCD analysis of HERA data}},
  \href{http://dx.doi.org/10.1140/epjc/s10052-015-3710-4}{{\em Eur. Phys. J. C}
  {\bf 75} (2015)~no.~12 580} [\href{http://arXiv.org/abs/1506.06042}{{\tt
  arXiv:1506.06042 [hep-ex]}}].

\bibitem{H1:2013ktq}
{\bf H1} collaboration, V.~Andreev {\em et.~al.}, {\it {Measurement of
  inclusive $e p$ cross sections at high $Q^2$ at $\sqrt s =$ 225 and 252 GeV
  and of the longitudinal proton structure function $F_L$ at HERA}},
  \href{http://dx.doi.org/10.1140/epjc/s10052-014-2814-6}{{\em Eur. Phys. J. C}
  {\bf 74} (2014)~no.~4 2814} [\href{http://arXiv.org/abs/1312.4821}{{\tt
  arXiv:1312.4821 [hep-ex]}}].

\bibitem{ZEUS:2014thn}
{\bf ZEUS} collaboration, H.~Abramowicz {\em et.~al.}, {\it {Deep inelastic
  cross-section measurements at large $y$ with the ZEUS detector at HERA}},
  \href{http://dx.doi.org/10.1103/PhysRevD.90.072002}{{\em Phys. Rev. D} {\bf
  90} (2014)~no.~7 072002} [\href{http://arXiv.org/abs/1404.6376}{{\tt
  arXiv:1404.6376 [hep-ex]}}].

\bibitem{Berge:1989hr}
J.~P. Berge {\em et.~al.}, {\it {A Measurement of Differential Cross-Sections
  and Nucleon Structure Functions in Charged Current Neutrino Interactions on
  Iron}},  \href{http://dx.doi.org/10.1007/BF01555493}{{\em Z. Phys. C} {\bf
  49} (1991) 187}.

\bibitem{CHORUS:2005cpn}
{\bf CHORUS} collaboration, G.~Onengut {\em et.~al.}, {\it {Measurement of
  nucleon structure functions in neutrino scattering}},
  \href{http://dx.doi.org/10.1016/j.physletb.2005.10.062}{{\em Phys. Lett. B}
  {\bf 632} (2006) 65}.

\bibitem{NuTeV:2005wsg}
{\bf NuTeV} collaboration, M.~Tzanov {\em et.~al.}, {\it {Precise measurement
  of neutrino and anti-neutrino differential cross sections}},
  \href{http://dx.doi.org/10.1103/PhysRevD.74.012008}{{\em Phys. Rev. D} {\bf
  74} (2006) 012008} [\href{http://arXiv.org/abs/hep-ex/0509010}{{\tt
  arXiv:hep-ex/0509010}}].

\bibitem{NuTeV:2001dfo}
{\bf NuTeV} collaboration, M.~Goncharov {\em et.~al.}, {\it {Precise
  Measurement of Dimuon Production Cross-Sections in $\nu_{\mu}$ Fe and
  $\bar{\nu}_{\mu}$ Fe Deep Inelastic Scattering at the Tevatron.}},
  \href{http://dx.doi.org/10.1103/PhysRevD.64.112006}{{\em Phys. Rev. D} {\bf
  64} (2001) 112006} [\href{http://arXiv.org/abs/hep-ex/0102049}{{\tt
  arXiv:hep-ex/0102049}}].

\bibitem{NOMAD:2013hbk}
{\bf NOMAD} collaboration, O.~Samoylov {\em et.~al.}, {\it {A Precision
  Measurement of Charm Dimuon Production in Neutrino Interactions from the
  NOMAD Experiment}},
  \href{http://dx.doi.org/10.1016/j.nuclphysb.2013.08.021}{{\em Nucl. Phys. B}
  {\bf 876} (2013) 339} [\href{http://arXiv.org/abs/1308.4750}{{\tt
  arXiv:1308.4750 [hep-ex]}}].

\bibitem{AbdulKhalek:2021gbh}
R.~Abdul~Khalek {\em et.~al.}, {\it {Science Requirements and Detector Concepts
  for the Electron-Ion Collider}: {EIC Yellow Report}},
  \href{http://dx.doi.org/10.1016/j.nuclphysa.2022.122447}{{\em Nucl. Phys. A}
  {\bf 1026} (2022) 122447} [\href{http://arXiv.org/abs/2103.05419}{{\tt
  arXiv:2103.05419 [physics.ins-det]}}].

\bibitem{LHeC:2020van}
{\bf LHeC, FCC-he Study Group} collaboration, P.~Agostini {\em et.~al.}, {\it
  {The Large Hadron-Electron Collider at the HL-LHC}},
  \href{http://dx.doi.org/10.1088/1361-6471/abf3ba}{{\em J. Phys. G} {\bf 48}
  (2021)~no.~11 110501} [\href{http://arXiv.org/abs/2007.14491}{{\tt
  arXiv:2007.14491 [hep-ex]}}].

\end{thebibliography}\endgroup

\end{document}